\UseRawInputEncoding 

\documentclass[]{article}

\usepackage{amsmath,amssymb}

\DeclareFontFamily{U}{musix}{}%
\DeclareFontShape{U}{musix}{m}{n}{%
  <-12>   musix11
  <12-15> musix13
  <15-18> musix16
  <18-23> musix20
  <23->   musix29
}{}%
\newcommand*\musix{\usefont{U}{musix}{m}{n}\selectfont}
\DeclareTextFontCommand{\textmusix}{\musix}

\usepackage[dvipdfmx]{graphicx}

\usepackage[top=30truemm,bottom=30truemm,left=25truemm,right=25truemm]{geometry}

\usepackage{amsthm}
\usepackage{bm}
\usepackage{color}
\usepackage{here}
\usepackage{comment}

\theoremstyle{plain}

\theoremstyle{definition}

\theoremstyle{remark}

\begin{document}

\title{General Theory of Music by Icosahedron 2: 
\\
\Large
Analysis of musical pieces by the exceptional musical icosahedra}

\author{Yusuke Imai}
\date{%
\small
    Graduate School of Engineering Science, Osaka University, Toyonaka, Osaka 560-8531, Japan\\%
    \today
}

\maketitle

\small

\begin{center}
CONTACT: 93imaiyusuke@gmail.com
\end{center}

\begin{abstract}
We propose a new way of analyzing musical pieces by using the exceptional musical icosahedra where all the major/minor triads are represented by golden triangles or golden gnomons.

First, we introduce a concept of the golden neighborhood that characterizes golden triangles/gnomons that neighbor a given golden triangle or gnomon. Then, we investigate a relation between the exceptional musical icosahedra and the neo-Riemannian theory, and find that the golden neighborhoods and the icosahedron symmetry relate any major/minor triad with any major/minor triad.

Second, we show how the exceptional musical icosahedra are applied to analyzing harmonies constructed by four or more tones. We introduce two concepts, golden decomposition and golden singular. The golden decomposition is a decomposition of a given harmony into the minimum number of harmonies constructing the given harmony and represented by the golden figure (a golden triangle, a golden gnomon, or a golden rectangle). A harmony is golden singular if and only if the harmony does not have golden decompositions. We show results of the golden analysis (analysis by the golden decomposition) of the tertian seventh chords and the mystic chord. While the dominant seventh chord is the only tertian seventh chord that is golden singular in the type ${\rm 1}^*$ and the type ${\rm 4}^*$ exceptional musical icosahedron, the half-diminished seventh chord is the only tertian seventh chord that is golden singular in the type ${\rm 2}^*$ and the type ${\rm 3}^*$ exceptional musical icosahedron. Also, from the viewpoint of the golden decomposition, the minor major seventh chord can be regarded as a generalized major-minor dual of the augmented major seventh chord with some transpositions and interchanges. In addition, all the harmonies constructed by five or more tones are not golden singular.

Last, we apply the golden analysis to the famous prelude in C major composed by Johann Sebastian Bach (BWV 846). We found 7 combinations of the golden figures (a golden triangle, a golden gnomon, two golden triangles, two golden gnomons, a golden rectangle, a golden rectangle and a golden triangle, a golden rectangle and a golden gnomon) on the type ${\rm 2}^*$ or the type ${\rm 3}^*$ exceptional musical icosahedron dually represent all the measures of the BWV 846.


\end{abstract}

\newpage
\section{Introduction}
The analysis of things is central to human curiosity. For example, physicists have analyzed physical phenomena and elegant laws of nature have been found: Newton's law, Maxwell's law, relativity theory, quantum theory, Hubble-Lemaˆitre law, etc. People have also been interested in the mind and various psychological concepts that capture mental motions have been analyzed such as stream of consciousness, emotion, habit, will, unconsciousness, free association, neurosis. In addition, analyzing characteristics of societies is one of the major matters of concern by people and various kinds of social systems have been analyzed such as capitalism, socialism, communism, democracy, fascism, liberalism, libertarianism, etc.

Similar to the above examples, analysis of music should be regarded as an important subject because music may be a universal culture and a musical piece reflect the era and the place where the musical piece is created, and music has been connected with human emotions in various situations such as rituals, concerts, daily life, etc.

\begin{figure}[b]
\centering
{%
\resizebox*{14cm}{!}{\includegraphics{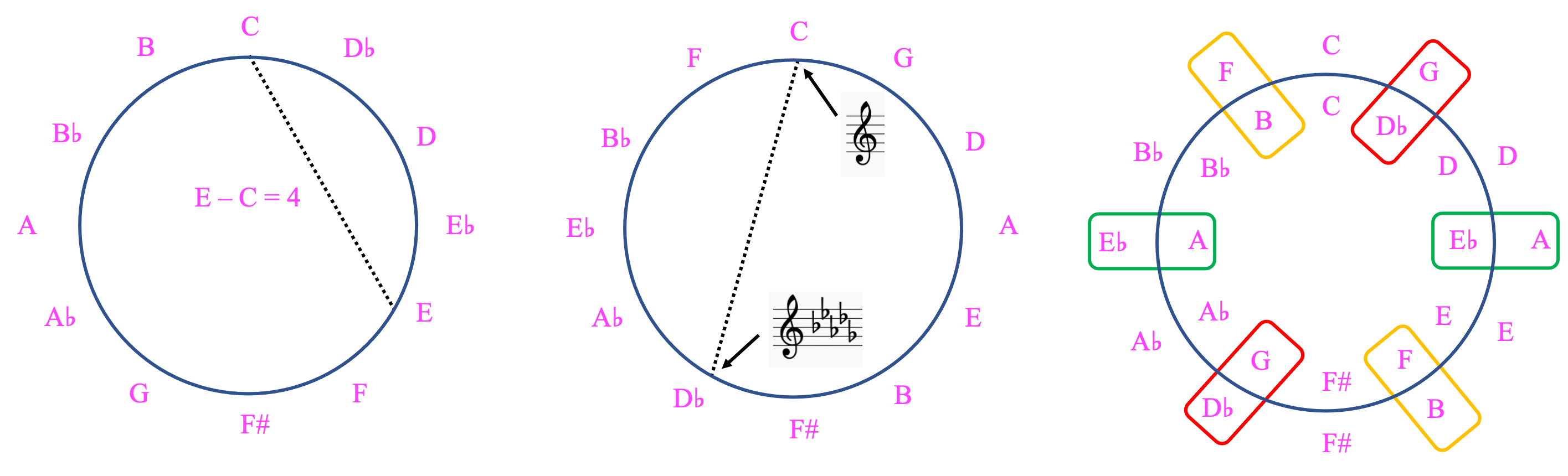}}}\hspace{5pt}
\caption{[Left] the chromatic circle (graph geodesic of $C$ and $E$ is 4 and $E$ is higher than $C$ by 4 semitones), [middle] the circle of fifths (graph geodesic of $C$ and $D\flat$ is 5 and the key of $D\flat$ has 5 more flats than the key of $C$), [right] comparison between the chromatic circle and the circle of fifths.} \label{chromatic_fifth_circle}
\end{figure}

\begin{figure}[t]
\centering
{%
\resizebox*{10cm}{!}{\includegraphics{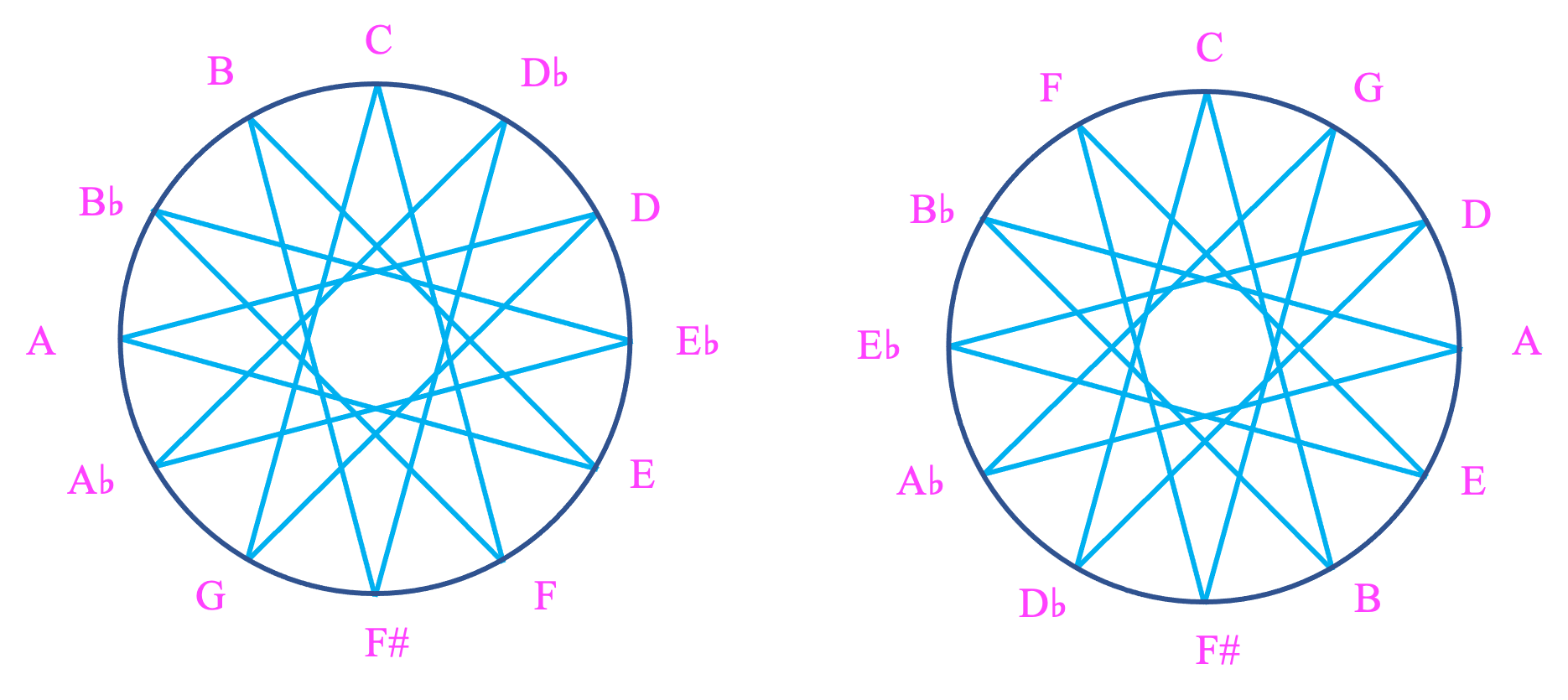}}}\hspace{5pt}
\caption{[Left] the twelve-pointed star shows the circle of fifths on the chromatic circle, [right] the twelve-pointed star shows the chromatic circle on the circle of fifths.} \label{chromatic_fifth_duality}
\end{figure}

Natural ways to analyze a relation among 12 tones are using the chromatic circle and the circle of fifths (McCartin, 2012). While the chromatic circle (Fig. 1, left) shows closeness of 12 tones, the circle of fifths (Fig. 1, middle) shows closeness of 12 keys. If an arrangement of 12 tones on the circle is equivalent to the reversed arrangement, those two circles are the only circles that satisfy the following two conditions: (i) intervals of neighboring two tones is constant, (ii) 12 tones are arranged without duplication. Then, one can show a duality between the chromatic circle and the circle of fifths by drawing the circle of fifths on the chromatic circle and the chromatic circle on the circle of fifths (Fig. 2). The figure of the cirlce of fifth on the chromatic circle is same as the figure of the chromatic cirlce on the circle of fifths. Therefore, for a theory of the circle of fifths, one can create a theory of the chromatic circle that is the dual theory of the circle of fifths. These circles are useful to deal with modulation (change of a key) and twelve-note composition (Headlam, Hasegawa, Lansky, and Perle, 2013).

\begin{figure}[H]
\centering
{%
\resizebox*{8cm}{!}{\includegraphics{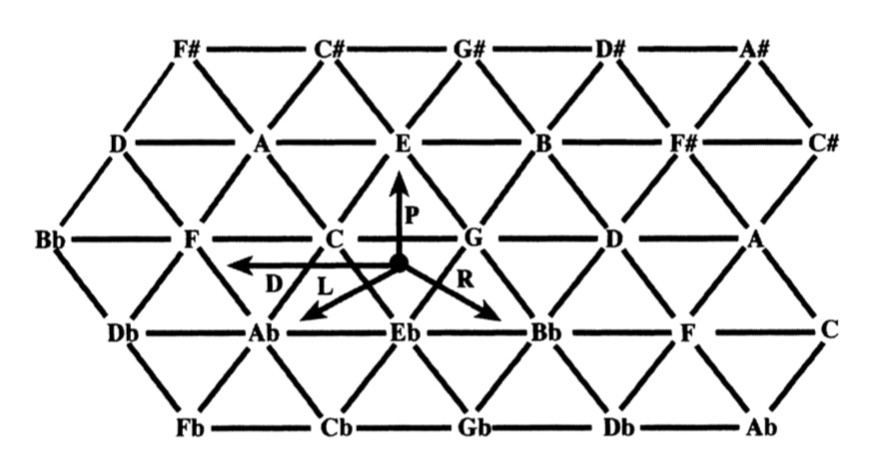}}}\hspace{5pt}
\caption{The \emph{Tonnetz} and three fundamental operations, $P$, $R$, $L$ and the transformation $D$ in the neo-Riemannian theory (Cohn, 1998).} \label{tonnetz}
\end{figure}

\begin{figure}[H]
\centering
{%
\resizebox*{14cm}{!}{\includegraphics{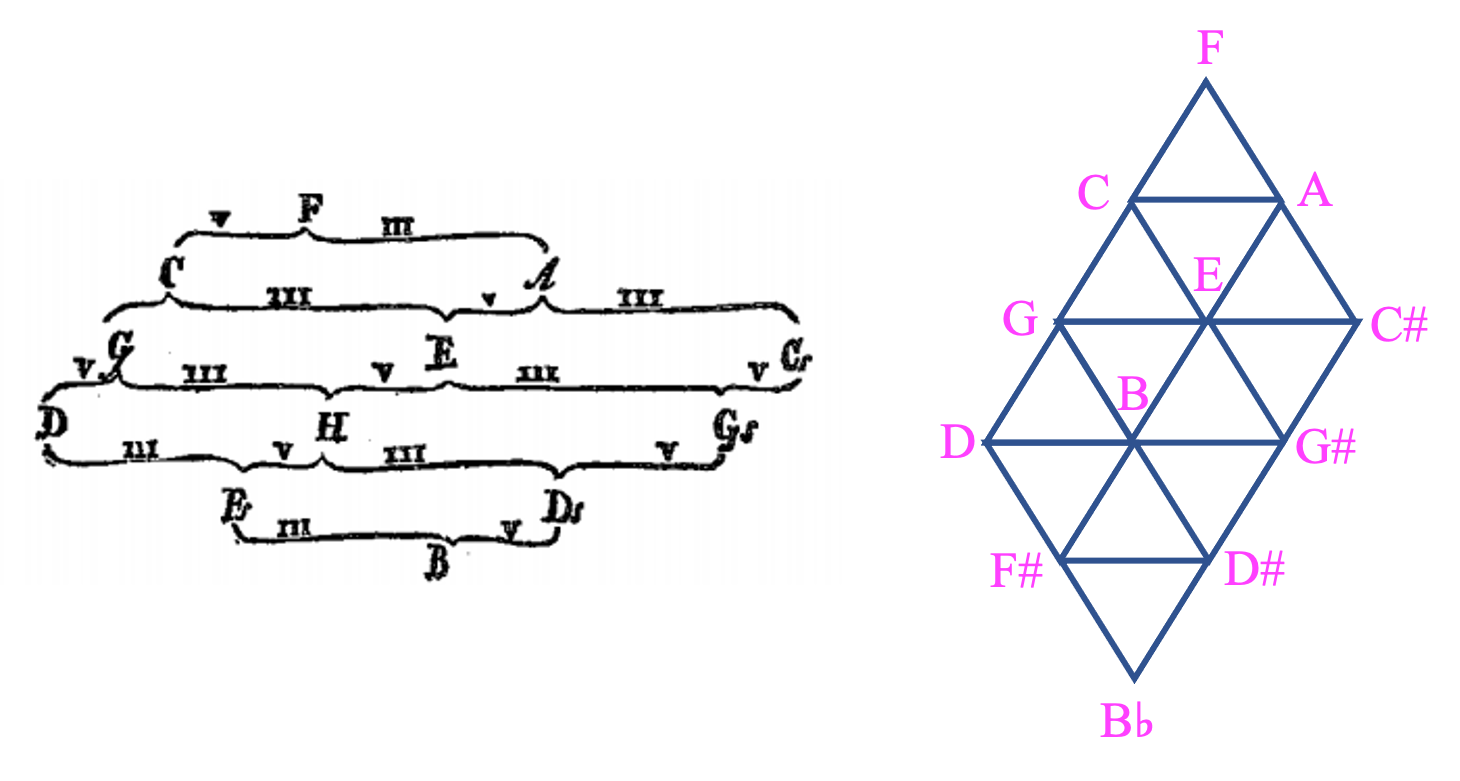}}}\hspace{5pt}
\caption{Euler's version of the \emph{Tonnetz} (Euler, 1739) and the modified version of it by using a triangular lattice.} \label{tonnetz_euler}
\end{figure}

However, these graphs do not give us enough methods to analyze harmonies. Fundamental theories requiring a small number of assumptions, in terms of the Occam's razor, for analyzing harmonies have been investigated through Riemannian theory, Neo-Riemannian theory, and its generalizations. Karl Wilhelm Julius Hugo Riemann proposed the so-called Riemannian theory that involves two classes of transformations: \emph{Schritt} and \emph{Wechsel} (Klumpenhouwer, 1994). \emph{Schritt} transforms the $X$ major scale into the $Y$ major scale and \emph{Wechsel} transforms the $X$ major/minor scale into the $X$ minor/major scale. The neo-Riemannian theory has been proposed by David Lewin, Brian Hyer, Richard Cohn, Henry Klumpenhouwer, and the theory can deal with ``chromatic music that is triadic but not altogether tonally unified" (Cohn, 1998). Cohn remarked ``such characteristics are primarily identified with music of Wagner, Liszt, and subsequent generations, but are also represented by some passages from Mozart, Schubert, and other pre-1850 composers". The neo-Riemannian theory is based on the following three transformations: $P$ (Parallel), $R$ (Relative), and $L$ (Leading-Tone Exchange). $P$ transforms the $X$ major/minor scale into the $X$ minor/major scale. $R$ transforms the $X$ major scale into the $X+9$ minor scale and the $X$ minor scale into the $X+3$ major scale (Note that $3+9=12$.) $L$ transforms the $X$ major scale into the $X+4$ minor scale and the $X$ minor scale into the $X+8$ major scale (Note that $4+8=12$). By combining these three transformations, one can analyze various musical pieces shown above. The neo-Riemannian theory can be characterized by simple geometry, \emph{Tonnetz} (Cohn, 1997). Figure \ref{tonnetz} shows three fundamental operations, $P$, $R$, $L$, and $D$ that transforms the $X$ major/minor triad into the $X+5$ minor/major triad in the \emph{Tonnetz}. Note that the rightward straight lines on the lattice show the sequence $X$, $X+7$, $X+14$, $\cdots$, and the up-right-ward straight lines show $X$, $X+4$, $X+8$, $\cdots$, and all the upward (downward) triangles in the \emph{Tonnetz} represent a major (minor) triad. The remarkable point is that in \emph{Tonnetz}, $P$, $R$, and $L$ transform a triangle into its adjacent triangles with respect to the edge. \emph{Tonnetz} was firstly proposed by celebrated Leonhard Euler (Euler, 1739). Figure \ref{tonnetz_euler} shows Euler's version of the \emph{Tonnetz}, and it includes the $F$ major triad, the $C$ major triad, the $A$ minor triad, the $A$ major triad, the $G$ major triad, the $E$ minor triad, $E$ major triad, the $C\sharp$ minor triad, the $B$ minor triad, the $B$ major triad, $G\sharp$ minor triad, and the $D\sharp$ minor triad.

\begin{figure}[H]
\centering
{%
\resizebox*{10cm}{!}{\includegraphics{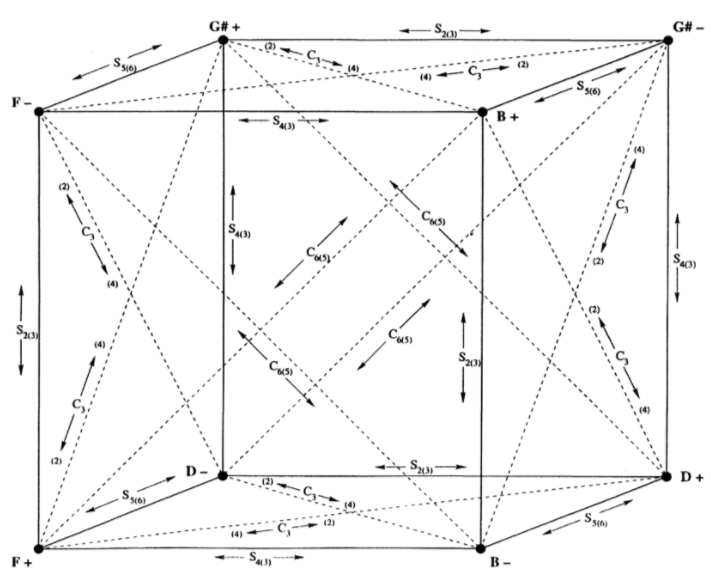}}}\hspace{5pt}
\caption{The 3d version of \emph{Tonnetz} by Childs. ``The cube involves three of the $S$ transforms, which form the solid edges, and the three $C$ transforms, which form the dotted-line diagonals on the faces" (Childs, 1998).} \label{3d_tonnetz}
\end{figure}

The neo-Riemannian theory has been used for analyzing various musical pieces and developed in various viewpoints. The neo-Riemannian theory has been used to analyze two pieces from Kurt\'ag's Kafka-Fragmente (Clough, 2002), pop-rock music (Capuzzo, 2004), jazz music (Briginshaw, 2012), musical pieces composed by Schubert (Rusch, 2013), film music (Lehman, 2014). An extension of the neo-Riemannian theory that deals with seventh codes (Fig.\,\ref{3d_tonnetz}) was proposed by Childs (Childs, 1998). In the paper by Childs, the dominant seventh code and the half-diminished seventh code are special, and this specialty has an important meaning in this paper. Embedding of the \emph{Tonnetz} to hypersphere in 4 dimensions (the Planet-4D model) was also proposed (Baroin, 2011). Furthermore, an extension of the \emph{Tonnetz} to N-tone equally tempered scales (for all N) and arbitrary triads was studied (Catanzaro, 2011). Amiot investigated a continuous version of \emph{Tonnetz} (the Torii of Phases) and its applications (Amiot, 2013). An extension of the \emph{Tonnetz} in terms of group extension was also presented (Popoff, 2013) and Popoff showed ``how group extensions can be used to build transformational models of time-spans and rhythms". In addition, Mohanty proposed a 5-dimensional \emph{Tonnetz}.

\begin{figure}[H]
\centering
{%
\resizebox*{14cm}{!}{\includegraphics{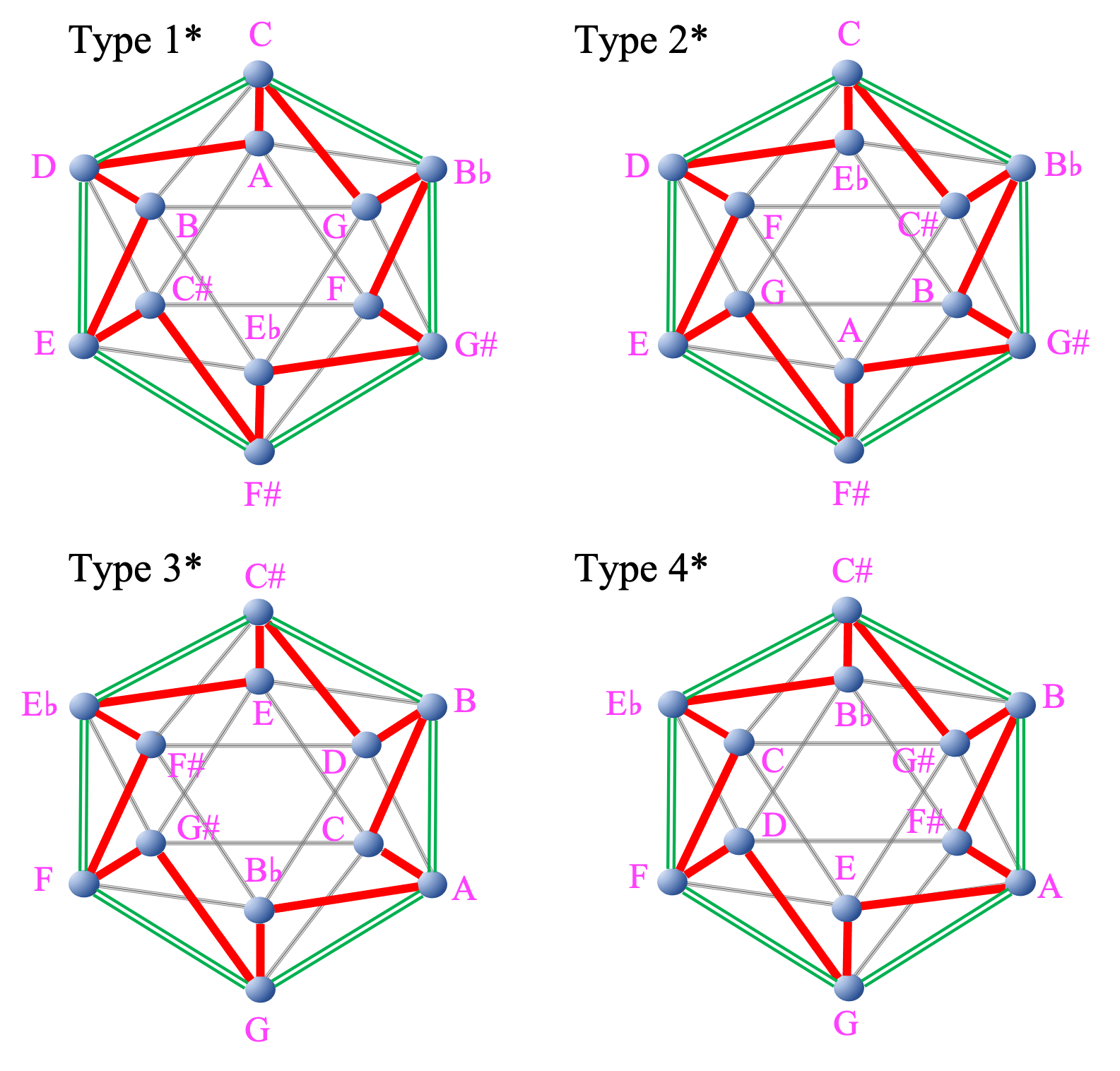}}}\hspace{5pt}
\caption{Four types of the exceptional musical icosahedra.} \label{emi}
\end{figure}

In our previous paper (Imai, Dellby Tanaka, 2021), we introduced a concept of the musical icosahedron that is the regular icosahedra each of whose vertices has one of the 12 tones without duplication, and we introduced some important musical icosahedra that connect various musical concepts: the chromatic/whole tone musical icosahedra, the Pythagorean/whole tone musical icosahedra, and the exceptional musical icosahedra. In this paper, we focus on the exceptional musical icosahedra (Fig.\,\ref{emi}) because they have the special characteristic, the golden major minor self-duality (all the major triads and minor triads are represented by some golden triangles and golden gnomons in all the types of the exceptional musical icosahedra). Therefore, the exceptional musical icosahedra can be regarded as one of the most appropriate models to analyze harmonies.

This paper is organized as follows. In Sec.\,II, we show a relation between the exceptional musical icosahedra and the neo-Riemannian theory. In Sec.\,III, we propose harmony analysis by the exceptional musical icosahedra. In Sec.\,IV, we perform the harmony analysis by the exceptional musical icosahedra to the BWV 846.

\newpage
\section{Exceptional Musical Icosahedra and Neo-Riemannian Theory}
In this section, we show a relation between the exceptional musical icosahedra and the neo-Riemannian theory. First, we propose a concept of the golden neighborhood. Then, we show how the golden neighborhoods relate with the fundamental transformations in the Neo-Riemannian theory: $P$, $R$, and $L$. Also, we show a relation of any major/minor triad to any major/minor triad by the golden neighborhood and the icosahedron symmetry.

\subsection{Golden Neighborhood}
\begin{figure}[H]
\centering
{%
\resizebox*{10cm}{!}{\includegraphics{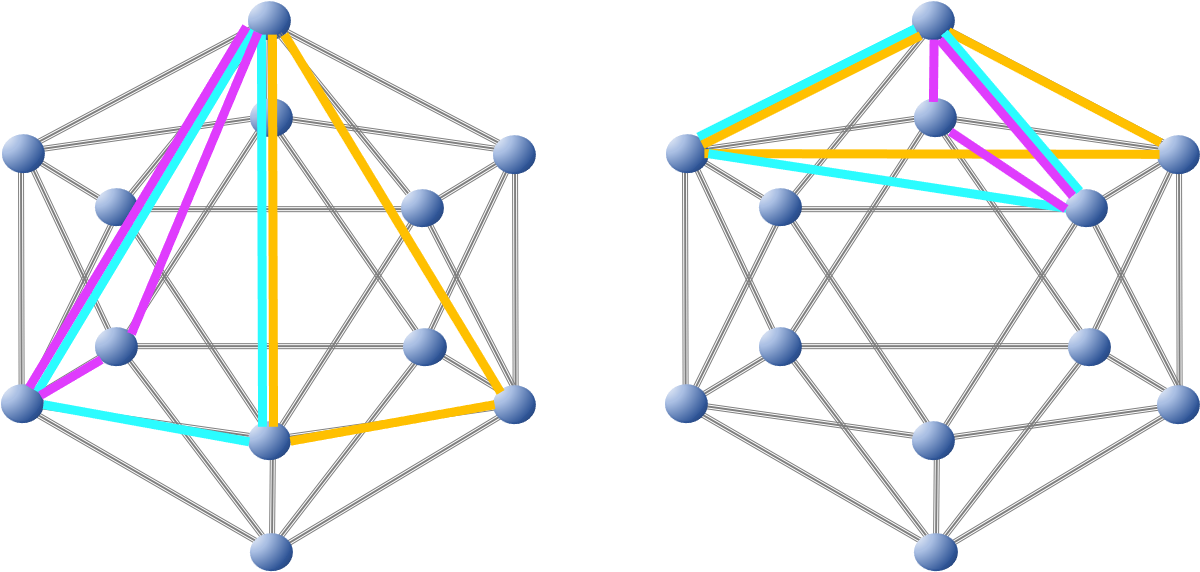}}}\hspace{5pt}
\caption{[Left] a golden triangle (cyan) and its golden-neighboring triangles (orange and purple), [right] a golden gnomon (cyan) and its golden-neighboring gnomons (orange and purple).} \label{golden_dual_tg}
\end{figure}

The exceptional musical icosahedra have the following fundamental property characterizing golden triangles/gnomons that neighbor a given golden triangle or gnomon.
\\
\\
\indent
[{\bf Golden Neighborhood 1}] For a given golden triangle on the regular icosahedron, there exist only two golden-neighboring triangles defined as golden triangles that share the apex and one of the long edges with the given golden triangle (Fig.\,\ref{golden_dual_tg}, left).
\\
\\
\indent
[{\bf Golden Neighborhood 2}] For a given golden gnomon on the regular icosahedron, there exist only two golden-neighboring gnomons defined as golden gnomons that share the apex and one of the short edges with the given golden triangle (Fig.\,\ref{golden_dual_tg}, right). 

\begin{figure}[H]
\centering
{%
\resizebox*{10cm}{!}{\includegraphics{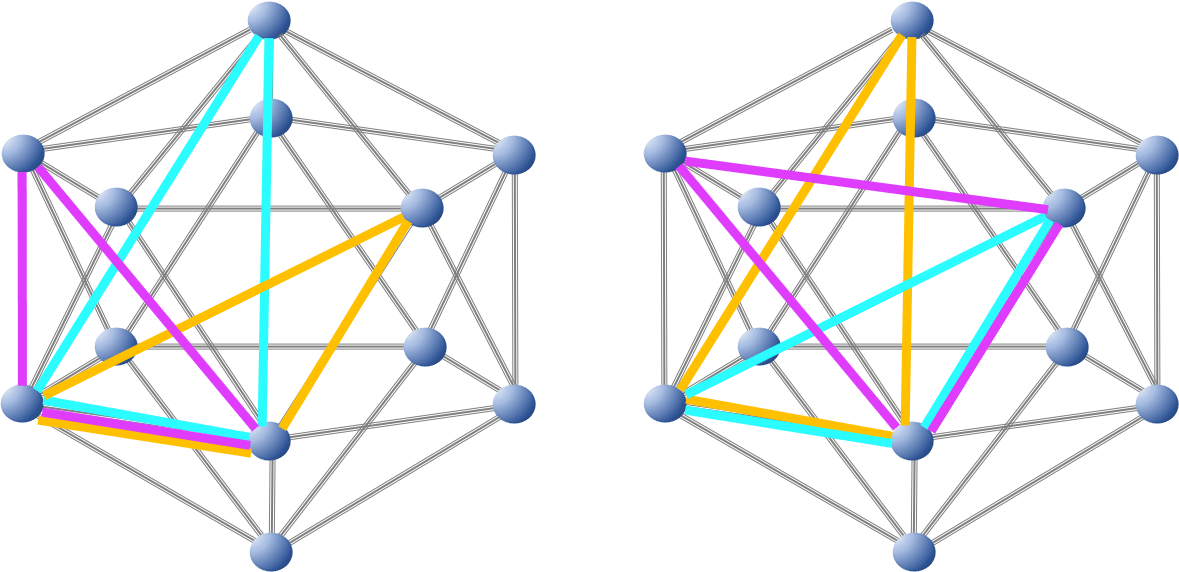}}}\hspace{5pt}
\caption{[Left] a golden triangle (cyan) and its golden-s-neighboring gnomons (orange and purple), [right] a golden gnomon (cyan) and its golden-s-neighboring triangles (orange and purple).} \label{golden_dual_gt}
\end{figure}

[{\bf Golden Neighborhood 3}]  For a given golden triangle on the regular icosahedron, there exist only two golden-s-neighboring gnomons defined as golden gnomons that share one of the short edges with the given golden triangle and that exist on the regular pentagon on which the given golden triangle exists (Fig.\,\ref{golden_dual_gt}, left).
\\
\\
\indent
[{\bf Golden Neighborhood 4}] For a given golden gnomon on the regular icosahedron, there exist only two golden-s-neighboring triangles defined as golden triangles that share the short edge with the given golden triangle and that exist on the regular pentagon on which the given golden gnomon exists (Fig.\,\ref{golden_dual_gt}, right).

\begin{figure}[H]
\centering
{%
\resizebox*{10cm}{!}{\includegraphics{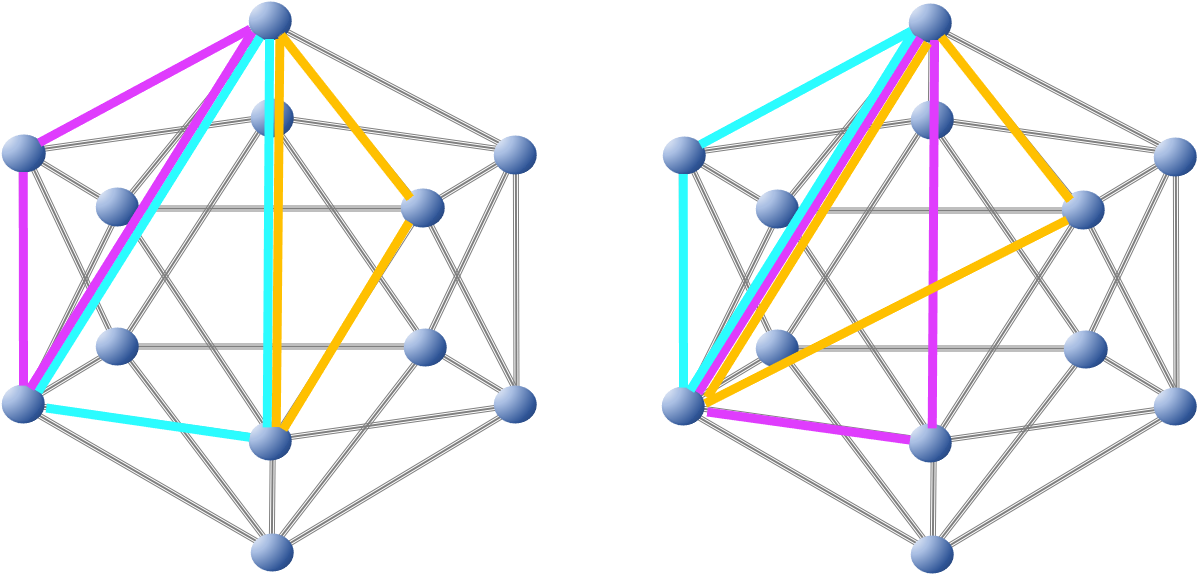}}}\hspace{5pt}
\caption{[Left] a golden triangle (cyan) and its golden-l-neighboring gnomons (orange and purple), [right] a golden gnomon (cyan) and its golden-l-neighboring triangles (orange and purple).} \label{golden_dual_gt2}
\end{figure}

[{\bf Golden Neighborhood 5}]  For a given golden triangle on the regular icosahedron, there exist only two golden-l-neighboring gnomons defined as golden gnomons that share the long edge with the given golden triangle and that exist on the regular pentagon on which the given golden triangle exists (Fig.\,\ref{golden_dual_gt2}, left).
\\
\\
\indent
[{\bf Golden Neighborhood 6}] For a given golden gnomon on the regular icosahedron, there exist only two golden-l-neighboring triangles defined as golden triangles that share one of the long edges with the given golden triangle and that exist on the regular pentagon on which the given golden gnomon exists (Fig.\,\ref{golden_dual_gt2}, right).

In the following subsections, we show how the golden neighborhoods characterize the fundamental operations in the neo-Riemannian theory: $P$, $R$, $L$. Because the exceptional musical icosahedra have the hexagon-icosahedron symmetry (an operation of raising all the tones of a given scale by two semitones corresponds to a symmetry transformation of the regular icosahedron), it is enough to deal with $P$, $R$, $L$ for the $C$ major triad and $C\sharp$ triad and there inverse transformations.

\subsection{Type ${\rm 1^*}$}
In the type ${\rm 1^*}$ exceptional musical icosahedron, the $C$ major triad is represented by the golden triangle, the $C$ minor triad by the golden gnomon, the $C\sharp$ major triad by the golden gnomon, the $C\sharp$ minor by the golden triangle. Then, the two golden-s-neighboring gnomons of the golden triangle representing the $C$ major triad represent a minor triad and a triad that is not a major triad or a minor triad. One can obtain uniquely the minor triad ($C$ minor triad) from the $C$ major triad. Therefore, a transformation of a golden triangle representing the $C$ major triad into the minor triad represented uniquely by one of the two golden-s-neighboring gnomons of the golden triangle representing the $C$ major triad is equivalent to the $P$-transformation for the $C$ major triad. The transformation of the $C$ minor triad into the $C$ major triad is also uniquely determined by one of the two golden-s-neighboring triangles of a golden gnomon representing the $C$ minor triad.

Also, $R$-transformation for the $C$ major triad is equivalent to a transformation of a golden triangle representing the $C$ major triad into the minor triad represented uniquely by one of the golden-neighboring triangles of a golden triangle representing the $C$ major triad. The transformation of the $A$ minor triad into the $C$ major triad is also uniquely determined by one of the two golden-neighboring triangles of a golden triangle representing the $A$ minor triad.

Furthermore, $L$-transformation for the $C$ major triad is equivalent to the three-fold rotation after the $P$-transformation for a golden triangle representing the $C$ major triad.

In addition, a transformation of a golden triangle representing the $C\sharp$ major triad into the  minor triad represented uniquely by one of the two golden-l-neighboring gnomons of the golden triangle representing the $C\sharp$ major triad is equivalent to the $P$-transformation for the $C\sharp$ major triad. The transformation of the $C\sharp$ minor triad into the $C\sharp$ major triad is also uniquely determined by one of the two golden-l-neighboring triangles of a golden gnomon representing the $C\sharp$ minor triad.

Also, $R$-transformation for the $C\sharp$ major triad is equivalent to a transformation of a golden gnomon representing the $C\sharp$ major triad into the minor triad represented uniquely by one of the golden-neighboring gnomons of the golden triangle representing the $C\sharp$ major triad. The transformation of the $B\flat$ minor triad into the $C\sharp$ major triad is also uniquely determined by one of the two golden-neighboring gnomons of the golden gnomon representing the $B\flat$ minor triad.

Furthermore, $L$-transformation for the $C\sharp$ major triad is equivalent to the three-fold rotation after the $P$-transformation for a golden gnomon representing the $C\sharp$ major triad.

\begin{figure}[H]
\centering
{%
\resizebox*{7cm}{!}{\includegraphics{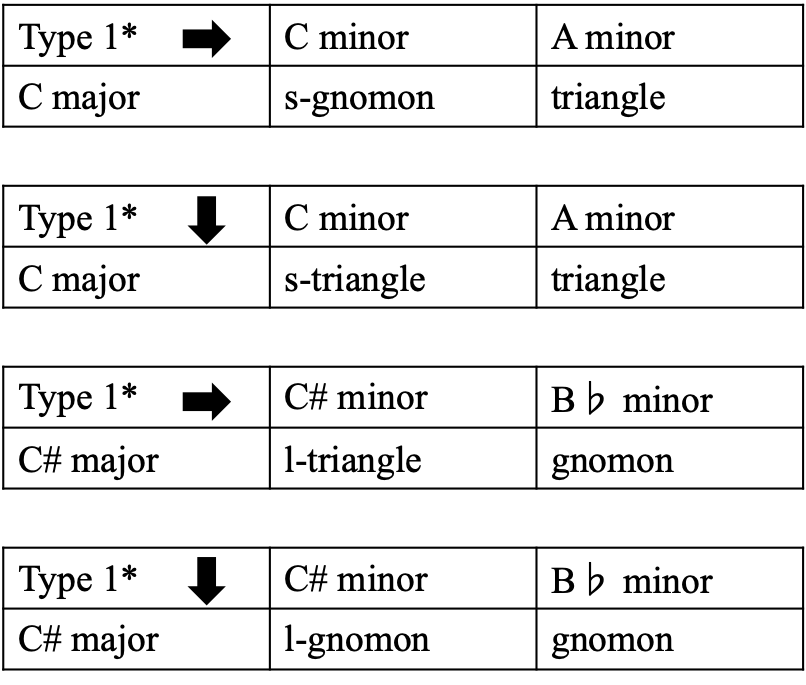}}}\hspace{5pt}
\caption{A summary of some golden neighborhoods in the type ${\rm 1^*}$ exceptional musical icosahedron corresponding to the $P$ and $R$ in the neo-Riemannian theory. The term ``triangle/gnomon" means golden-neighboring triangle/gnomon. The term s(l)-gnomon/triangle is an abbreviation for golden-s(l)-neighboring gnomon/triangle.} \label{type1_n1}
\end{figure}

\begin{figure}[H]
\centering
{%
\resizebox*{16cm}{!}{\includegraphics{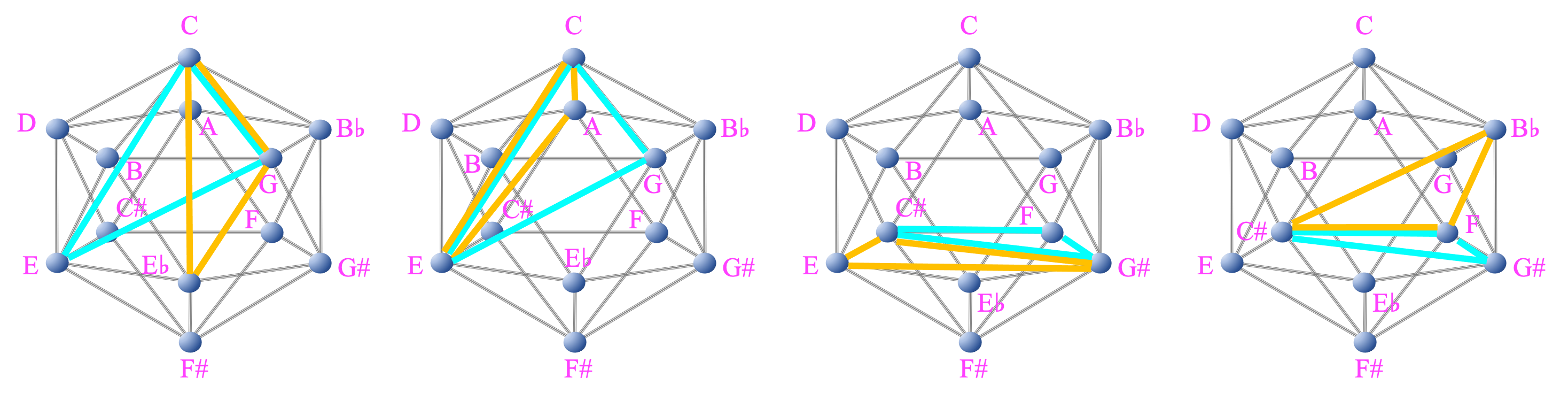}}}\hspace{5pt}
\caption{The $C$ major triad (cyan) and the $C$ minor triad (orange), the $C$ major triad (cyan) and the $A$ minor triad (orange), the $C\sharp$ major triad (cyan) and the $C\sharp$ minor triad (orange), and the $C\sharp$ major triad (cyan) and the $B\flat$ minor triad (orange) in the type ${\rm 1^*}$ exceptional musical icosahedron.} \label{type1_n2}
\end{figure}

These results are summarized in Fig.\,\ref{type1_n1} and Fig.\,\ref{type1_n2}.

\subsection{Type ${\rm 2^*}$}

\begin{figure}[H]
\centering
{%
\resizebox*{7cm}{!}{\includegraphics{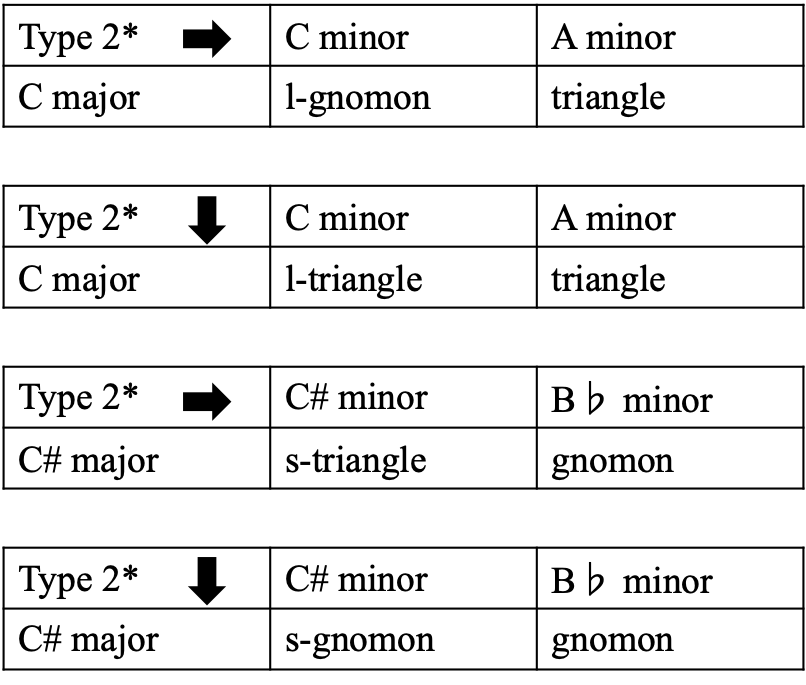}}}\hspace{5pt}
\caption{A summary of some golden neighborhoods in the type ${\rm 2^*}$ exceptional musical icosahedron corresponding to the $P$ and $R$ in the neo-Riemannian theory.} \label{type2_n1}
\end{figure}

\begin{figure}[H]
\centering
{%
\resizebox*{16cm}{!}{\includegraphics{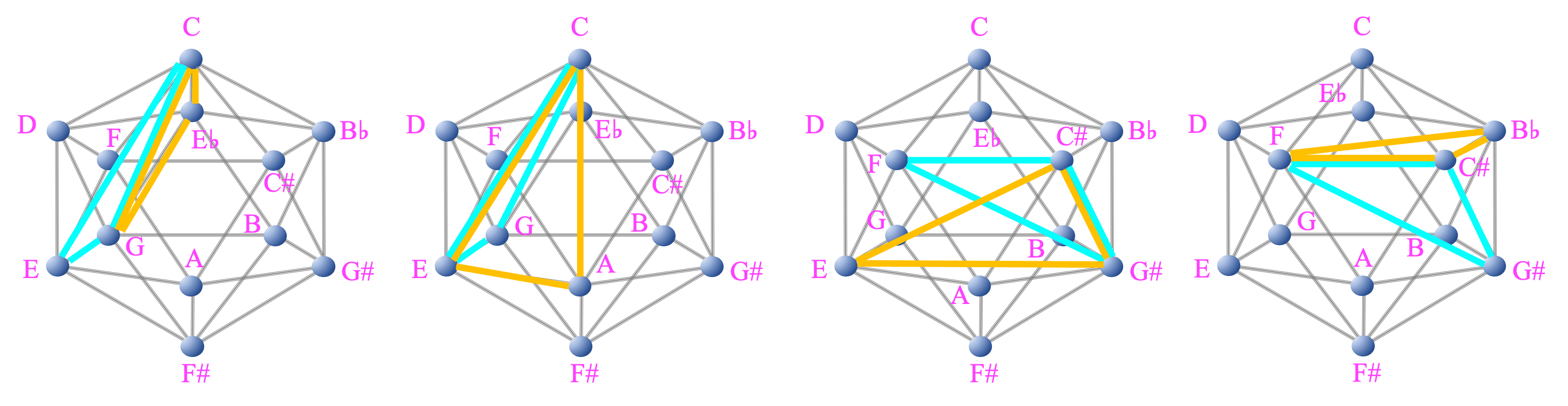}}}\hspace{5pt}
\caption{The $C$ major triad (cyan) and the $C$ minor triad (orange), the $C$ major triad (cyan) and the $A$ minor triad (orange), the $C\sharp$ major triad (cyan) and the $C\sharp$ minor triad (orange), and the $C\sharp$ major triad (cyan) and the $B\flat$ minor triad (orange) in the type ${\rm 2^*}$ exceptional musical icosahedron.} \label{type2_n2}
\end{figure}

Similar to the previous subsection, one has the equivalence between fundamental relations in the neo-Riemannian theory and the golden neighborhoods for the type ${\rm 2^*}$ in Fig.\,\ref{type2_n1} and Fig.\,\ref{type2_n2}.

\subsection{Type ${\rm 3^*}$}

\begin{figure}[H]
\centering
{%
\resizebox*{7cm}{!}{\includegraphics{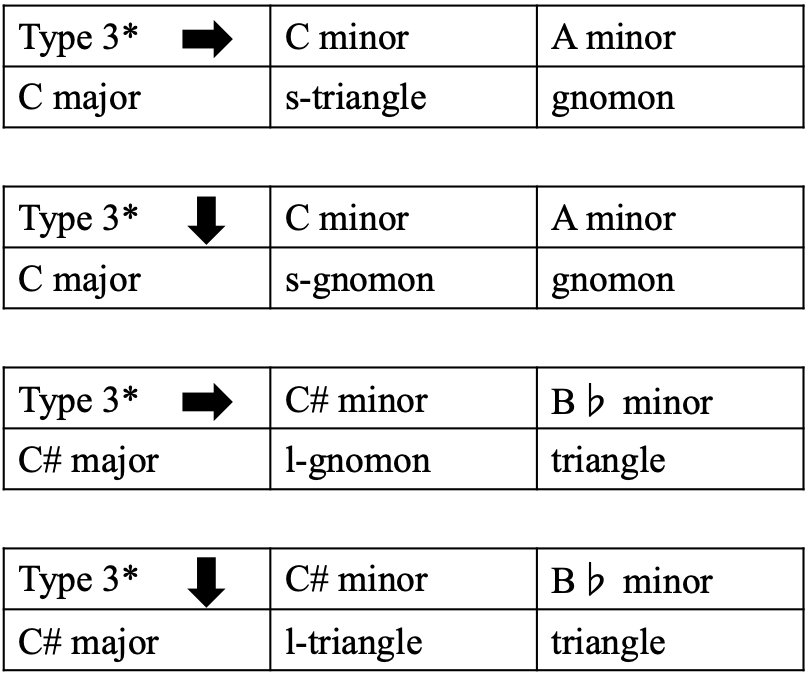}}}\hspace{5pt}
\caption{A summary of some golden neighborhoods in the type ${\rm 3^*}$ exceptional musical icosahedron corresponding to the $P$ and $R$ in the neo-Riemannian theory.} \label{type3_n1}
\end{figure}

\begin{figure}[H]
\centering
{%
\resizebox*{16cm}{!}{\includegraphics{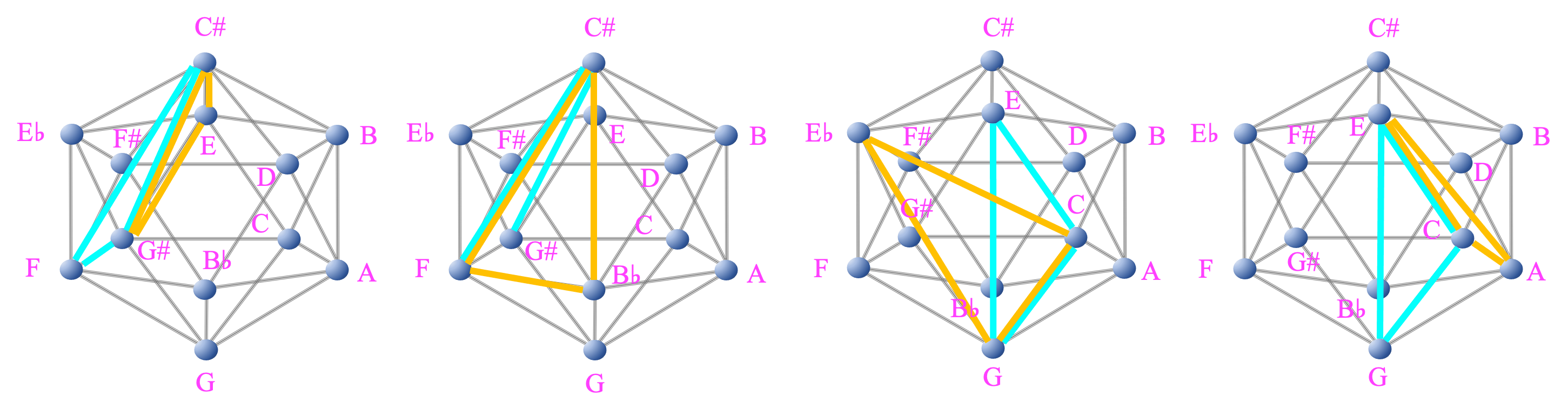}}}\hspace{5pt}
\caption{The $C\sharp$ major triad (cyan) and the $C\sharp$ minor triad (orange), the $C\sharp$ major triad (cyan) and the $B\flat$ minor triad (orange), the $C$ major triad (cyan) and the $C$ minor triad (orange), and the $C$ major triad (cyan) and the $A$ minor triad (orange) in the type ${\rm 3^*}$ exceptional musical icosahedron.} \label{type3_n2}
\end{figure}

Similar to the previous subsections, one has the equivalence between fundamental relations in the neo-Riemannian theory and the golden neighborhoods for the ${\rm 3^*}$ in Fig.\,\ref{type3_n1} and Fig.\,\ref{type3_n2}.

\subsection{Type ${\rm 4^*}$}

\begin{figure}[H]
\centering
{%
\resizebox*{7cm}{!}{\includegraphics{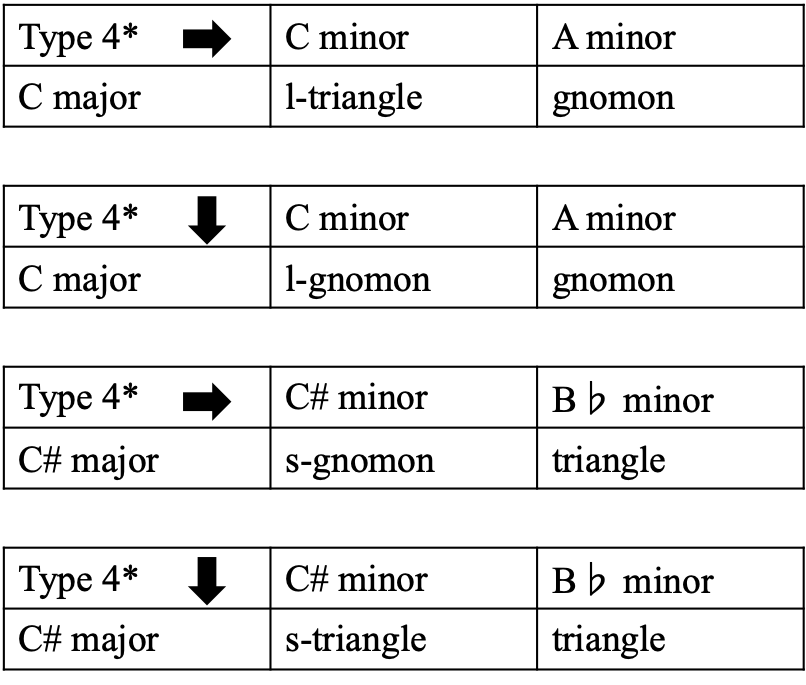}}}\hspace{5pt}
\caption{A summary of some golden neighborhoods in the type ${\rm 4^*}$ exceptional musical icosahedron corresponding to the $P$ and $R$ in the neo-Riemannian theory.} \label{type4_n1}
\end{figure}

\begin{figure}[H]
\centering
{%
\resizebox*{16cm}{!}{\includegraphics{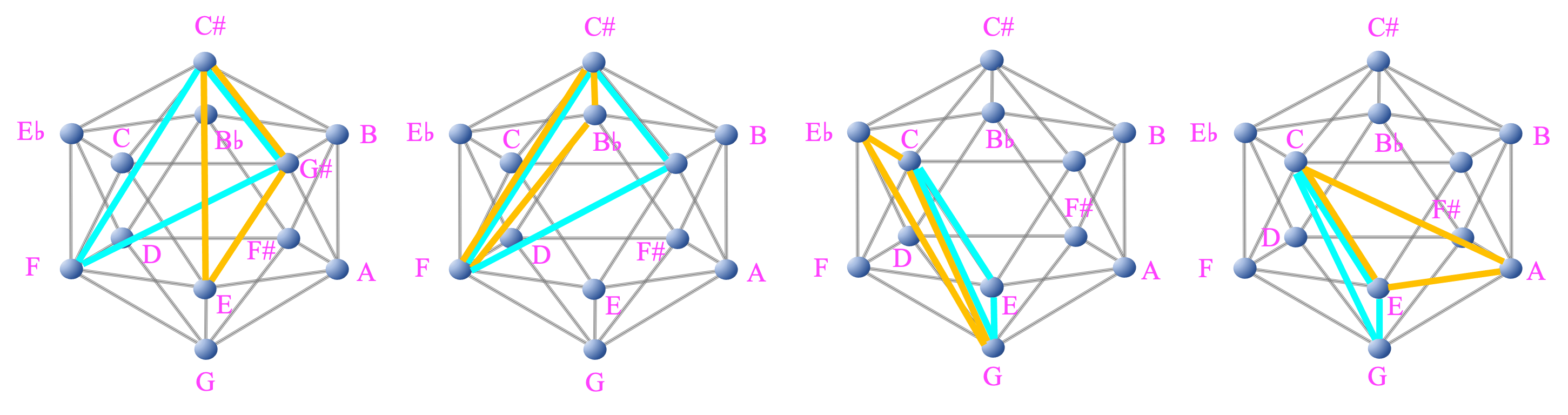}}}\hspace{5pt}
\caption{The $C\sharp$ major triad (cyan) and the $C\sharp$ minor triad (orange), the $C\sharp$ major triad (cyan) and the $B\flat$ minor triad (orange), the $C$ major triad (cyan) and the $C$ minor triad (orange), and the $C$ major triad (cyan) and the $A$ minor triad (orange) in the type ${\rm 4^*}$ exceptional musical icosahedron.} \label{type4_n2}
\end{figure}

Similar to the previous subsections, one has the equivalence between fundamental relations in the neo-Riemannian theory and the golden neighborhoods for the type ${\rm 4^*}$ in Fig.\,\ref{type4_n1} and Fig.\,\ref{type4_n2}.

\subsection{Relation between any major/minor triad and any major/minor triad}
In the previous subsections, we showed how the $C$/$C\sharp$ major triad is related to the $A$/$B\flat$ minor triad and $C$/$C\sharp$ minor triad. One can also relate the $A$/$B\flat$ major triad with the $C$/$C\sharp$ major triad because the $A$/$B\flat$ major triad can be related to the $A$/$B\flat$ minor triad in the same manner as $C$/$C\sharp$ major triad is related to the $C$/$C\sharp$ minor triad. Then, by considering the symmetry of the regular icosahedron, one can relate the $C$/$C\sharp$ major triad with the $C\sharp$/$C$ major triad, $E\flat$/$D$ major triad, $F$/$E$ major triad, $G$/$F\sharp$ major triad, $A$/$G\sharp$ major triad, and $B$/$B\flat$ major triad. Therefore, one can relate any major/minor triad to any major/minor triad by using the golden neighborhoods and the icosahedron symmetry on the exceptional musical icosahedra. This means that the $L$ transformation that is fundamental in the neo-Riemannian theory is not fundamental in this theory. Of course, there may exist other theories of musical icosahedra where the fundamental transformations are given by $P$, $R$, and $L$.


\newpage
\section{Harmony Analysis by Exceptional Musical Icosahedra}
In this section, we present how to analyze harmonies by the exceptional musical icosahedra. First, we prove a fundamental lemma of the exceptional musical icosahedra, and a structure of harmony analysis by the exceptional musical icosahedra is identified. Then, we obtain some properties of some harmonies constructed by four or more tones with the golden figures (a golden triangle, a golden gnomon, and a golden rectangle) on the regular icosahedron.

\subsection{A fundamental lemma of the exceptional musical icosahedra}
The exceptional musical icosahedra have a unique characteristic that the chromatic/whole tone musical icosahedra and the Pythagorean/whole tone musical icosahedra do not have. This characteristic is important to deal with harmony analysis by the exceptional musical icosahedra.
\\
\\
\indent
[{\bf Fundamental Lemma for the Exceptional Musical Icosahedra}]

\begin{figure}[H]
\centering
{%
\resizebox*{6cm}{!}{\includegraphics{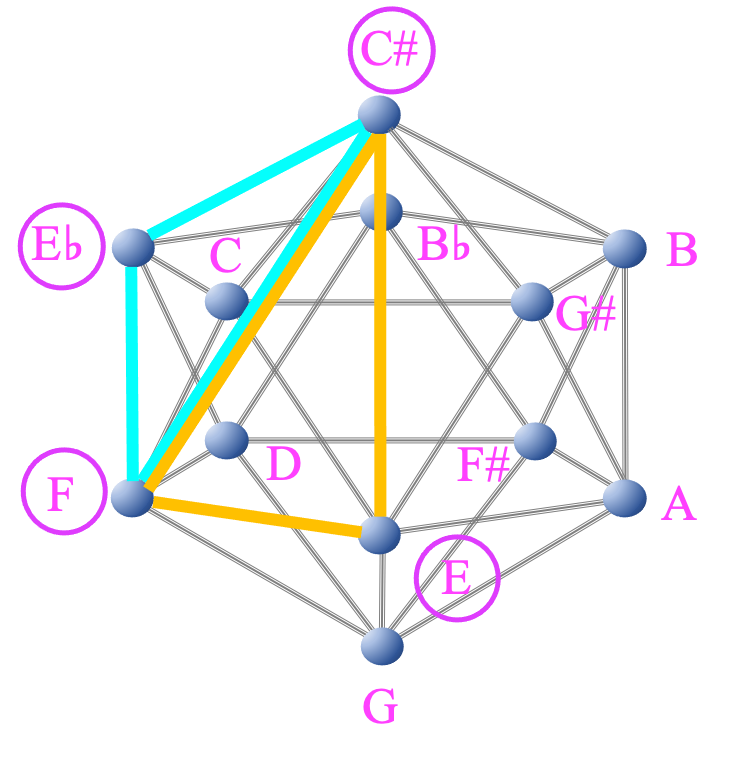}}}\hspace{5pt}
\caption{A harmony, $C\sharp$, $E\flat$, $E$, $F$, can be decomposed into a harmony, $C\sharp$, $E\flat$, $F$ (golden gnomon), and a harmony, $C\sharp$, $E$, $F$ (golden triangle).} \label{golden_example}
\end{figure}

On the exceptional musical icosahedra, if a triad, $X$, $Y$, $Z$, is represented by a golden triangle/gnomon, a triad, $X+1$, $Y+1$, $Z+1$, is represented by a golden gnomon/triangle.
\\
\\
\indent
Now, we create a harmony analysis by the exceptional musical icosahedra by decomposing a given harmony into the golden-base harmonies that are defined as harmonies constructing the given harmony and represented by the golden figures (Fig.\,\ref{golden_example}). We call such a harmony analysis golden analysis.

By combining the hexagon-icosahedron symmetry and the Fundamental Lemma for the Exceptional Musical Icosahedra, one can find that it is enough to show how harmonies based on $C$ are decomposed into some golden figures on the type ${\rm 1^*}$ and the type ${\rm 2^*}$ exceptional musical icosahedron. For any integer, $n$, if a harmony $X$, $Y$, $Z$, $W$, is represented by a golden rectangle on the type $n$ exceptional musical icosahedron, then, for any integers, $m$, $n'$, a harmony, $X+m$, $Y+m$, $Z+m$, $W+m$, is also represented by a golden rectangle on the type $n'$ exceptional musical icosahedron. If a harmony $X$, $Y$, $Z$, is represented by a golden triangle/gnomon on the type ${\rm 1^*}$ exceptional musical icosahedron, then, for any integer $m$, (i) a harmony, $X+2m$, $Y+2m$, $Z+2m$, is represented by a golden triangle/gnomon on the type ${\rm 1^*}$ exceptional musical icosahedron, and a harmony, $X+2m+1$, $Y+2m+1$, $Z+2m+1$, is represented by a golden gnomon/triangle on the type ${\rm 1^*}$ exceptional musical icosahedron, (ii) a harmony, $X+2m$, $Y+2m$, $Z+2m$, is represented by a golden gnomon/triangle on the type ${\rm 4^*}$ exceptional musical icosahedron, and a harmony, $X+2m+1$, $Y+2m+1$, $Z+2m+1$, is represented by a golden triangle/gnomon on the type ${\rm 4^*}$ exceptional musical icosahedron. Also, if a harmony $X$, $Y$, $Z$, is represented by a golden triangle/gnomon on the type ${\rm 2^*}$ exceptional musical icosahedron, then, for any integer $m$, (i) a harmony, $X+2m$, $Y+2m$, $Z+2m$, is represented by a golden triangle/gnomon on the type ${\rm 2^*}$ exceptional musical icosahedron, and a harmony, $X+2m+1$, $Y+2m+1$, $Z+2m+1$, is represented by a golden gnomon/triangle on the type ${\rm 2^*}$ exceptional musical icosahedron, (ii) a harmony, $X+2m$, $Y+2m$, $Z+2m$, is represented by a golden gnomon/triangle on the type ${\rm 3^*}$ exceptional musical icosahedron, and a harmony, $X+2m+1$, $Y+2m+1$, $Z+2m+1$, is represented by a golden triangle/gnomon on the type ${\rm 3^*}$ exceptional musical icosahedron.

\subsection{Golden decomposition and golden singular}
This subsection defines some fundamental concepts to deal with the golden analysis: golden decomposition and golden singular. We define golden decompositions of a harmony $A$ on the type ${\rm n^*}$ as decompositions of a figure representing $A$ by the minimum number of golden figures. Also, we define golden singular as follows on the type ${\rm n^*}$; a harmony $A$ is golden singular on the type ${\rm n^*}$ if and only if $A$ does not have golden decompositions.

\begin{figure}[H]
\centering
{%
\resizebox*{6cm}{!}{\includegraphics{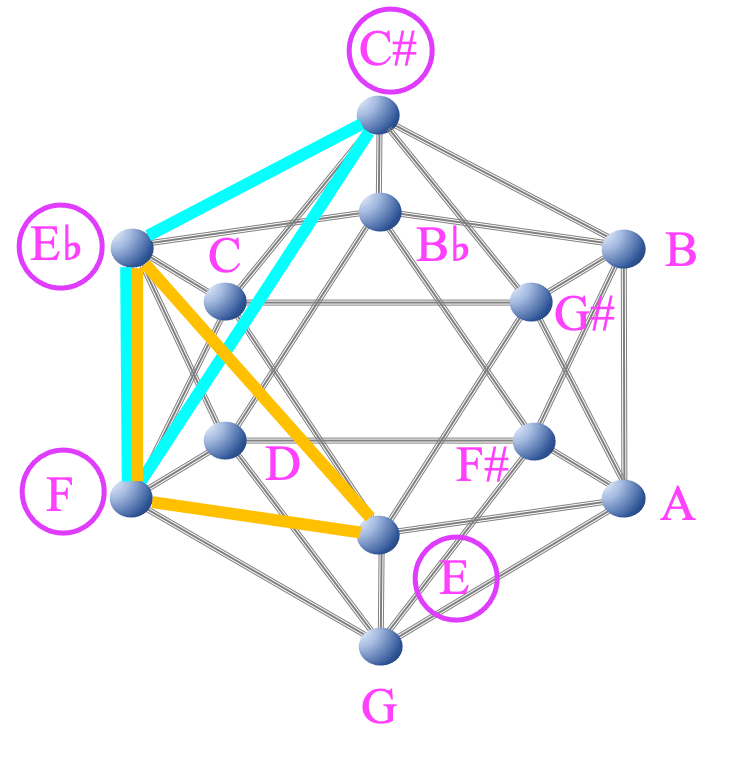}}}\hspace{5pt}
\caption{A harmony, $C\sharp$, $E\flat$, $E$, $F$, can be decomposed into a harmony, $C\sharp$, $E\flat$, $F$ (golden gnomon), and a harmony, $E\flat$, $E$, $F$ (golden gnomon).} \label{golden_example_2}
\end{figure}

Note that a harmony generally has many golden decompositions. For example, $C\sharp$, $E\flat$, $E$, $F$, shown in Fig.\,\ref{golden_example} has two golden decompositions (Fig.\,\ref{golden_example_2}).

\begin{figure}[H]
\centering
{%
\resizebox*{16cm}{!}{\includegraphics{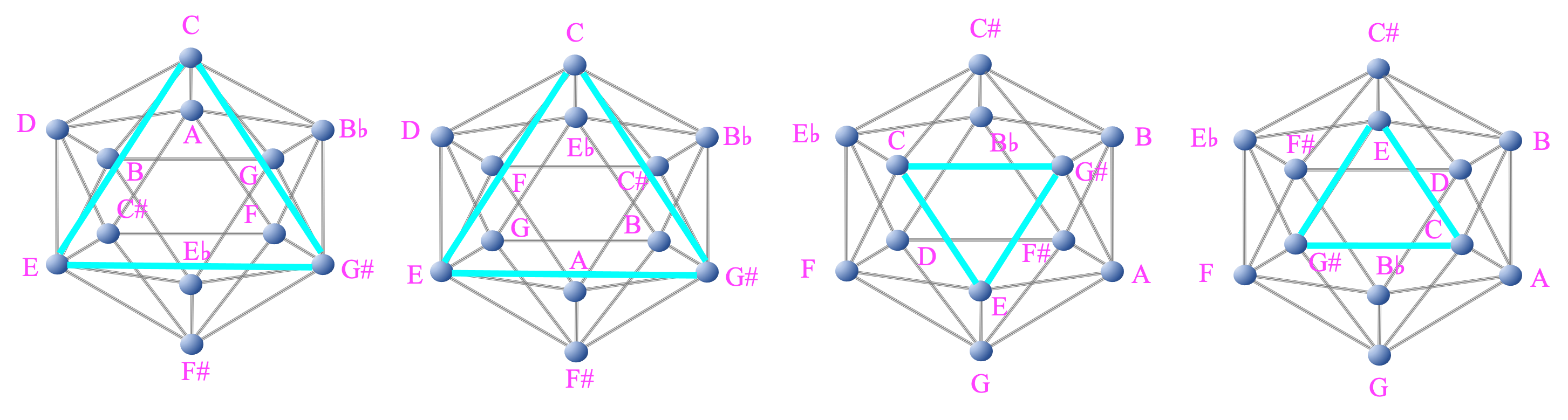}}}\hspace{5pt}
\caption{$C$, $E$, $G\sharp$ on the type ${\rm 1^*}$, the type ${\rm 2^*}$, the type ${\rm 3^*}$, the type ${\rm 4^*}$. Obviously, $C$, $E$, $G\sharp$ is golden singular on all the types of the exceptional musical icosahedra.} \label{aug}
\end{figure}

Note also that if a harmony $A$ is golden singular on the exceptional musical icosahedra, transpositions of $A$ are also golden singular because of the Fundamental Lemma for the Exceptional Musical Icosahedra. A simple example of a harmony that is golden singular on the exceptional musical icosahedra is $C$, $E$, $G\sharp$ because $C$, $E$, $G\sharp$ is represented by the equilateral triangle constructed by the edge of the regular icosahedron (Fig.\,\ref{aug}).

\subsection{Golden analysis for the seventh chords}
\begin{figure}[H]
\centering
{%
\resizebox*{16cm}{!}{\includegraphics{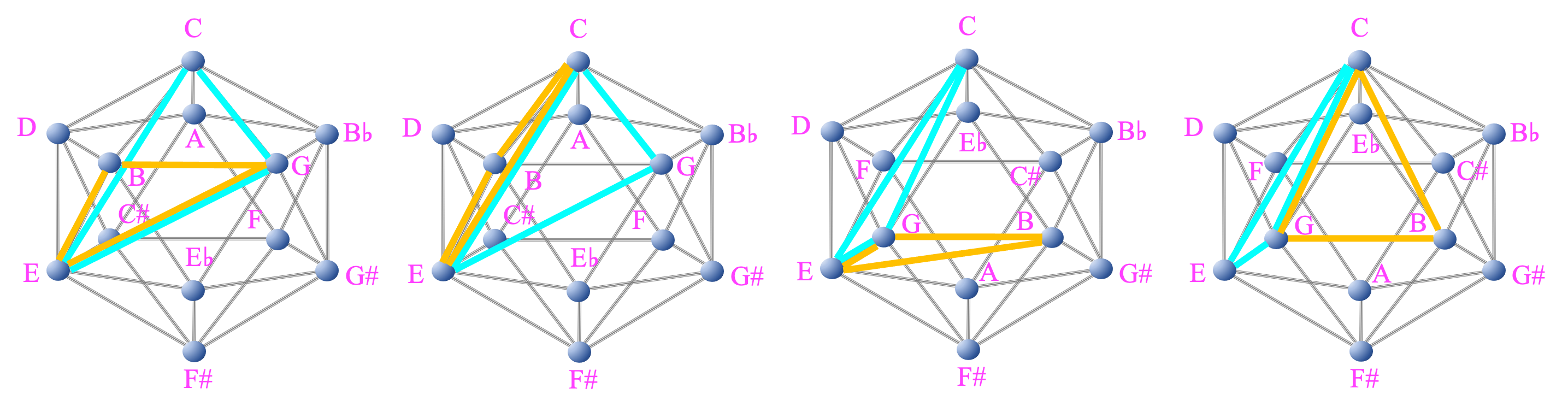}}}\hspace{5pt}
\caption{The golden analysis of $C$, $E$, $G$, $B$. There exist two golden decompostions of $C$, $E$, $G$, $B$, for the type ${\rm 1^*}$: $C$, $E$, $G$, $B$ = $C$, $E$, $G$ (gt) + $E$, $G$, $B$ (gg), $C$, $E$, $G$, $B$ = $C$, $E$, $G$ (gt) + $C$, $E$, $B$ (gg). There exist two golden decompostions of $C$, $E$, $G$, $B$, for the type ${\rm 2^*}$: $C$, $E$, $G$, $B$ = $C$, $E$, $G$ (gt) + $E$, $G$, $B$ (gg), $C$, $E$, $G$, $B$ = $C$, $E$, $G$ (gt) + $C$, $G$, $B$ (gt).}
\label{CEGB}
\end{figure}

\begin{figure}[H]
\centering
{%
\resizebox*{16cm}{!}{\includegraphics{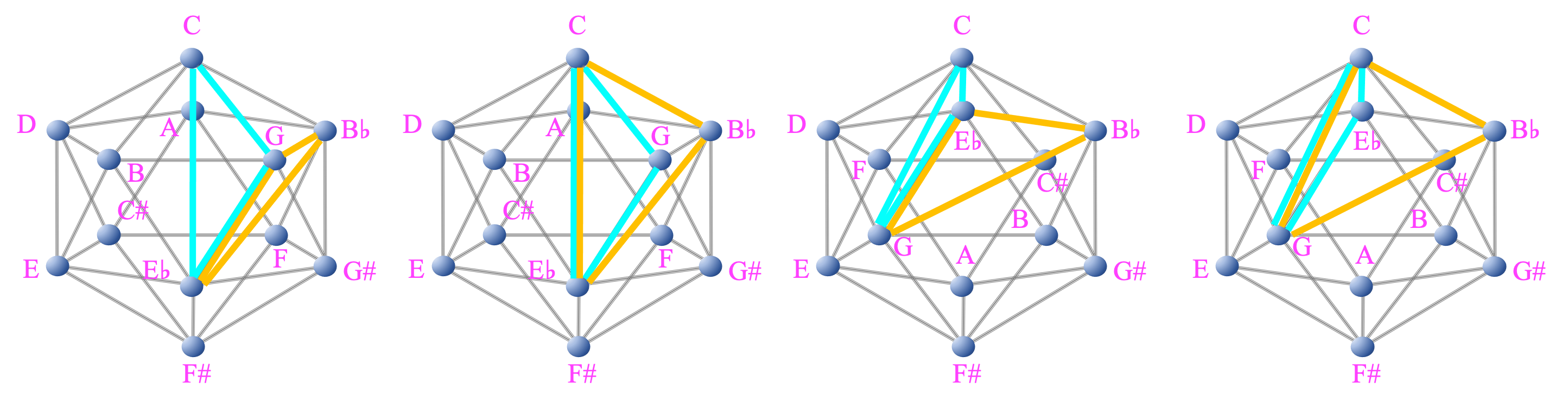}}}\hspace{5pt}
\caption{The golden analysis of $C$, $E\flat$, $G$, $B\flat$. There exist two golden decompostions of $C$, $E\flat$, $G$, $B\flat$, for the type ${\rm 1^*}$: $C$, $E\flat$, $G$, $B\flat$ = $C$, $E\flat$, $G$ (gg) + $E\flat$, $G$, $B\flat$ (gg), $C$, $E\flat$, $G$, $B\flat$ = $C$, $E\flat$, $G$ (gg) + $C$, $E\flat$, $B\flat$ (gt). There exist two golden decompostions of $C$, $E\flat$, $G$, $B\flat$, for the type ${\rm 2^*}$: $C$, $E\flat$, $G$, $B\flat$ = $C$, $E\flat$, $G$ (gg) + $E\flat$, $G$, $B$ (gg), $C$, $E\flat$, $G$, $B\flat$ = $C$, $E\flat$, $G$ (gg) + $C$, $G$, $B\flat$ (gt).}
\label{CEfGBf}
\end{figure}

\begin{figure}[H]
\centering
{%
\resizebox*{10cm}{!}{\includegraphics{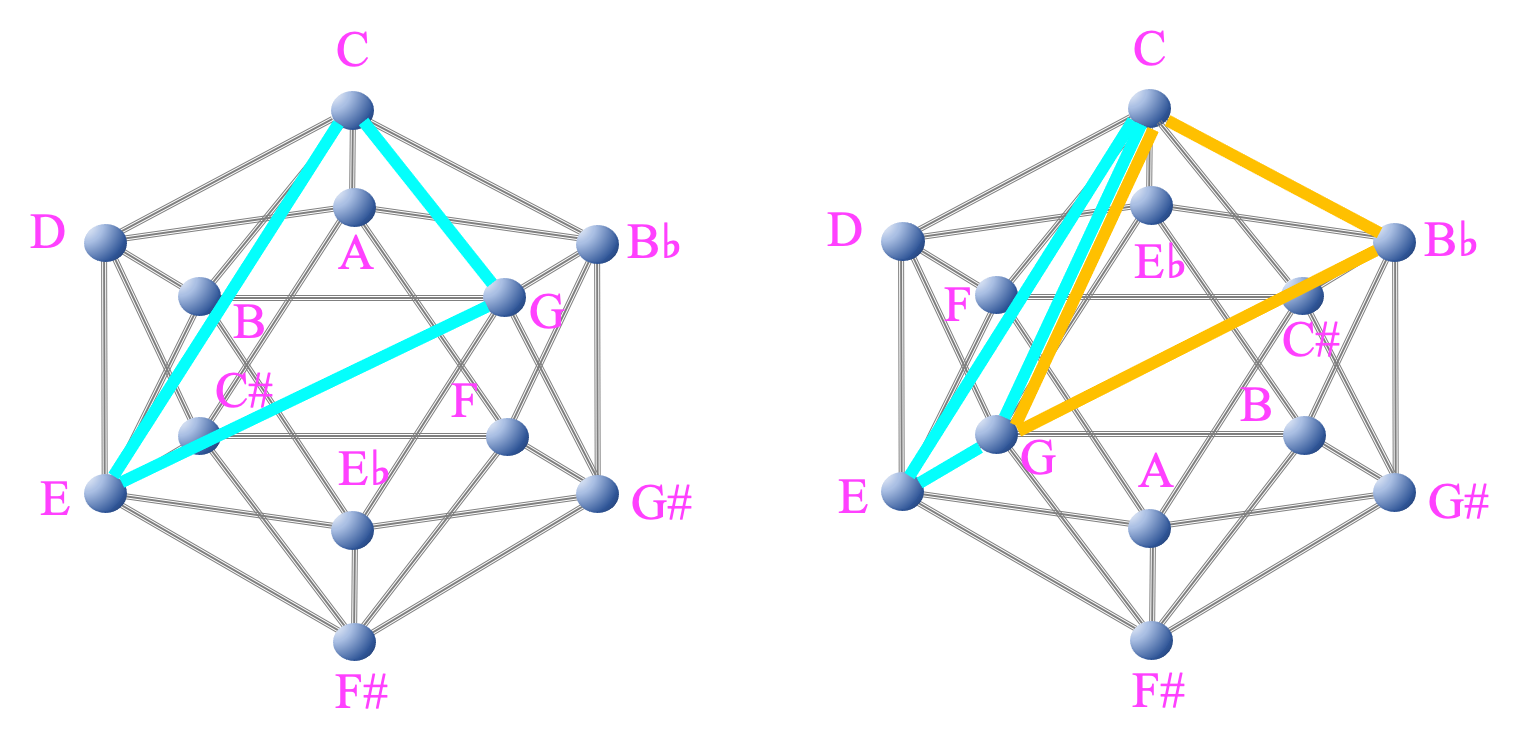}}}\hspace{5pt}
\caption{The golden analysis of $C$, $E$, $G$, $B\flat$. $C$, $E$, $G$, $B\flat$ is golden singular in the type ${\rm 1^*}$. There exists one golden decompostion of $C$, $E$, $G$, $B\flat$, for the type ${\rm 2^*}$: $C$, $E$, $G$, $B\flat$ = $C$, $E$, $G$ (gt) + $C$, $E$, $B\flat$ (gt).}
\label{CEGBf}
\end{figure}

\begin{figure}[H]
\centering
{%
\resizebox*{10cm}{!}{\includegraphics{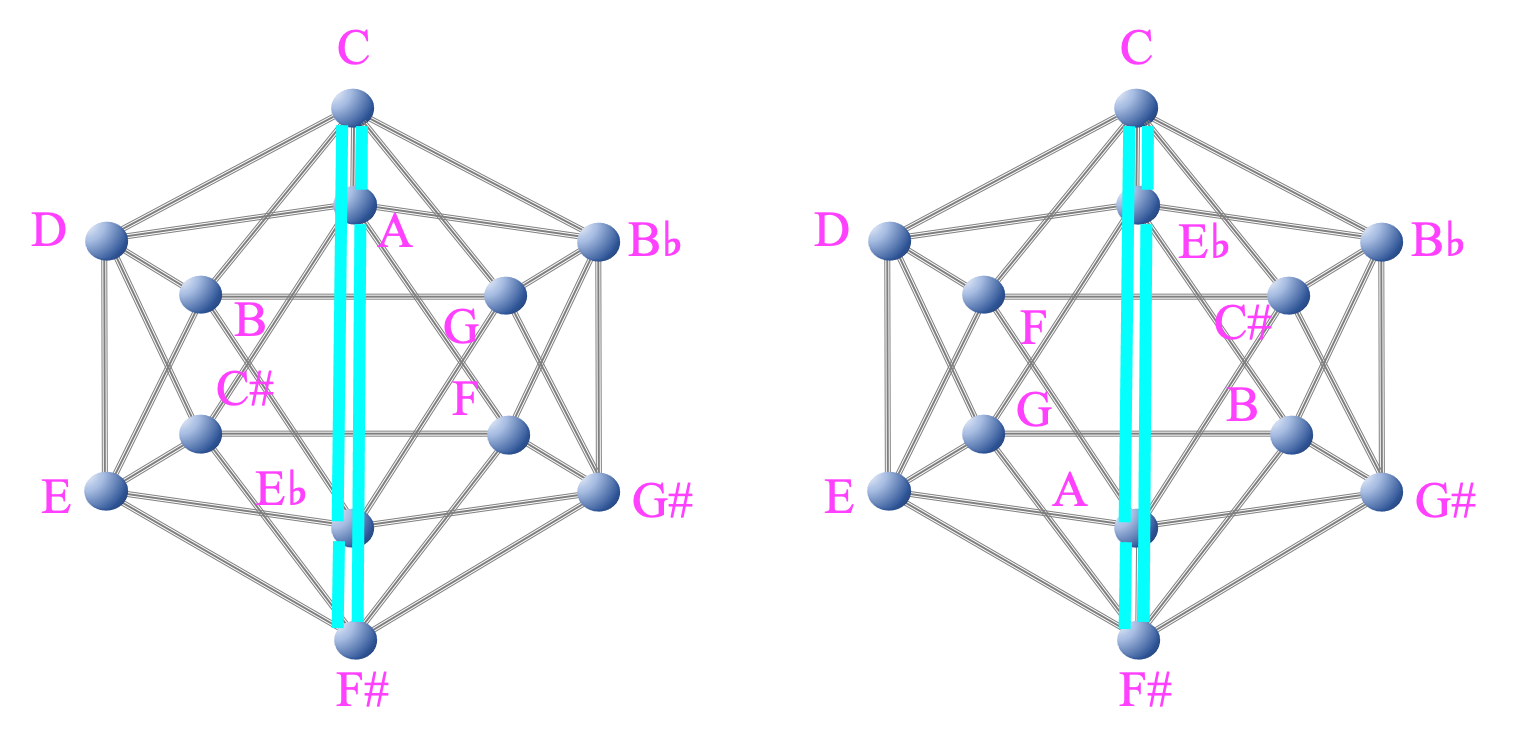}}}\hspace{5pt}
\caption{The golden analysis of $C$, $E\flat$, $G\flat$, $A$. There exists one golden decompostion of $C$, $E\flat$, $G\flat$, $A$, for the type ${\rm 1^*}$: $C$, $E\flat$, $G\flat$, $A$ (golden rectangle). There exists one golden decompostion of $C$, $E\flat$, $G\flat$, $A$, for the type ${\rm 2^*}$: $C$, $E\flat$, $G\flat$, $A$ (golden rectangle).}
\label{CEfGfA}
\end{figure}

\begin{figure}[H]
\centering
{%
\resizebox*{10cm}{!}{\includegraphics{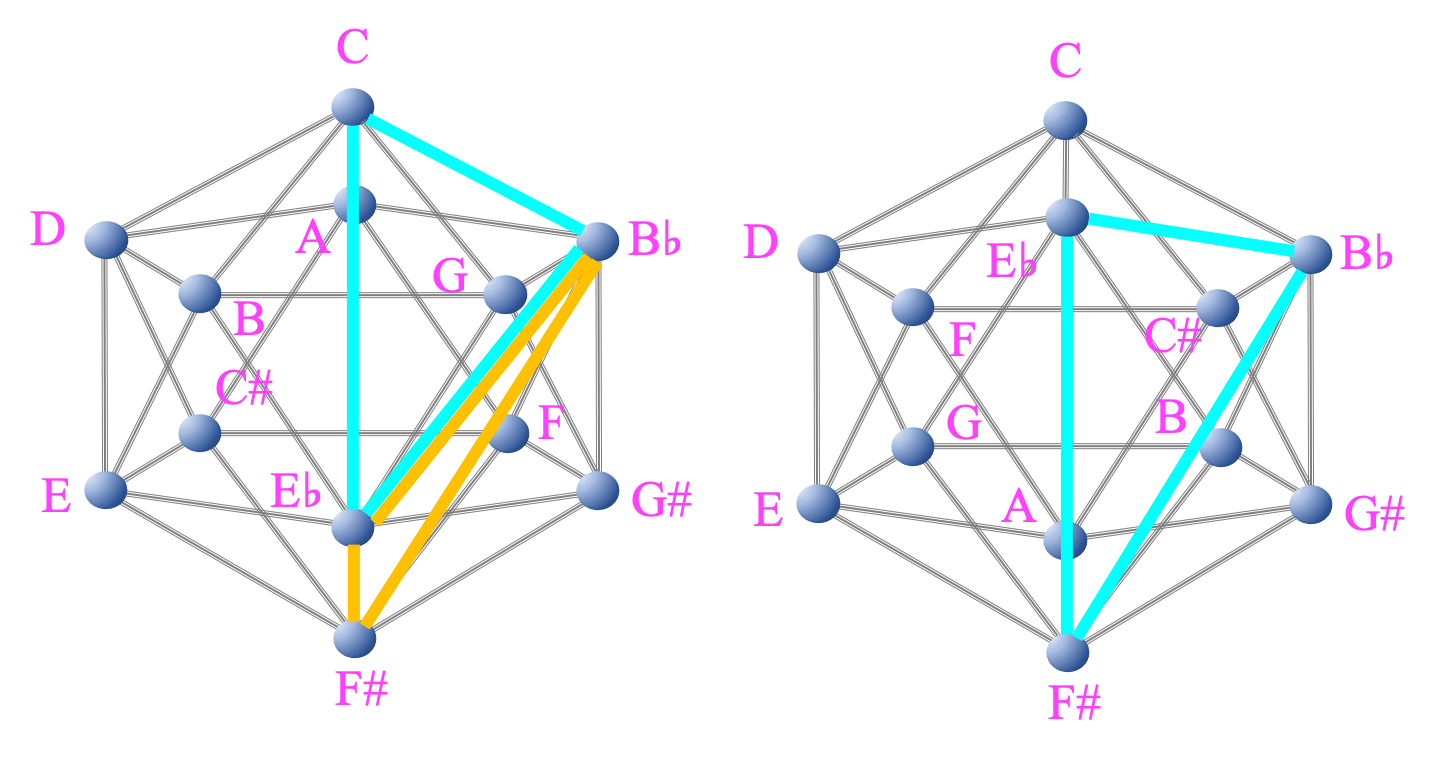}}}\hspace{5pt}
\caption{The golden analysis of $C$, $E\flat$, $G\flat$, $B\flat$. There exists one golden decompostion of $C$, $E\flat$, $G\flat$, $B\flat$, for the type ${\rm 1^*}$: $C$, $E\flat$, $G\flat$, $B\flat$ = $C$, $E\flat$, $B\flat$ (gt) + $E\flat$, $G\flat$, $B\flat$ (gt). $C$, $E\flat$, $G\flat$, $B\flat$ is golden singular in the type ${\rm 2^*}$.}
\label{CEfGfBf}
\end{figure}

\begin{figure}[H]
\centering
{%
\resizebox*{16cm}{!}{\includegraphics{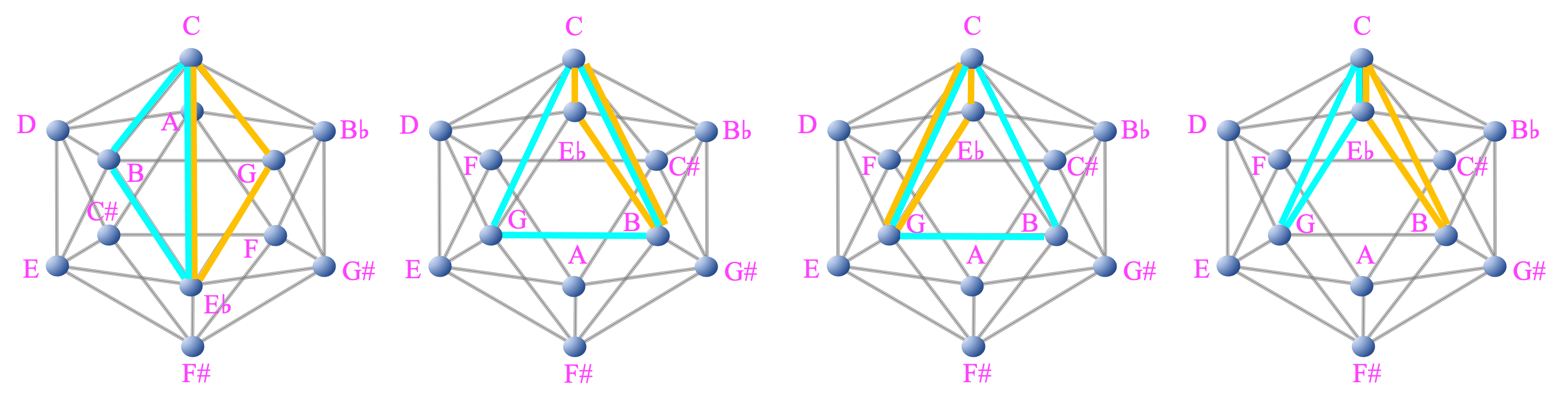}}}\hspace{5pt}
\caption{The golden analysis of $C$, $E\flat$, $G$, $B$. There exists one golden decompostion of $C$, $E\flat$, $G$, $B$, for the type ${\rm 1^*}$: $C$, $E\flat$, $G$, $B$ = $C$, $E\flat$, $G$ (gg) + $C$, $E\flat$, $B$ (gg). There exist three golden decompostions of $C$, $E\flat$, $G$, $B$, for the type ${\rm 2^*}$: $C$, $E\flat$, $G$, $B$ = $C$, $G$, $B$ (gt) + $C$, $E\flat$, $B$ (gg), $C$, $E\flat$, $G$, $B$ = $C$, $G$, $B$ (gt) + $C$, $E\flat$, $G$ (gg), $C$, $E\flat$, $G$, $B$ = $C$, $E\flat$, $G$ (gg) + $C$, $E\flat$, $B$ (gg).}
\label{CEfGB}
\end{figure}

\begin{figure}[H]
\centering
{%
\resizebox*{16cm}{!}{\includegraphics{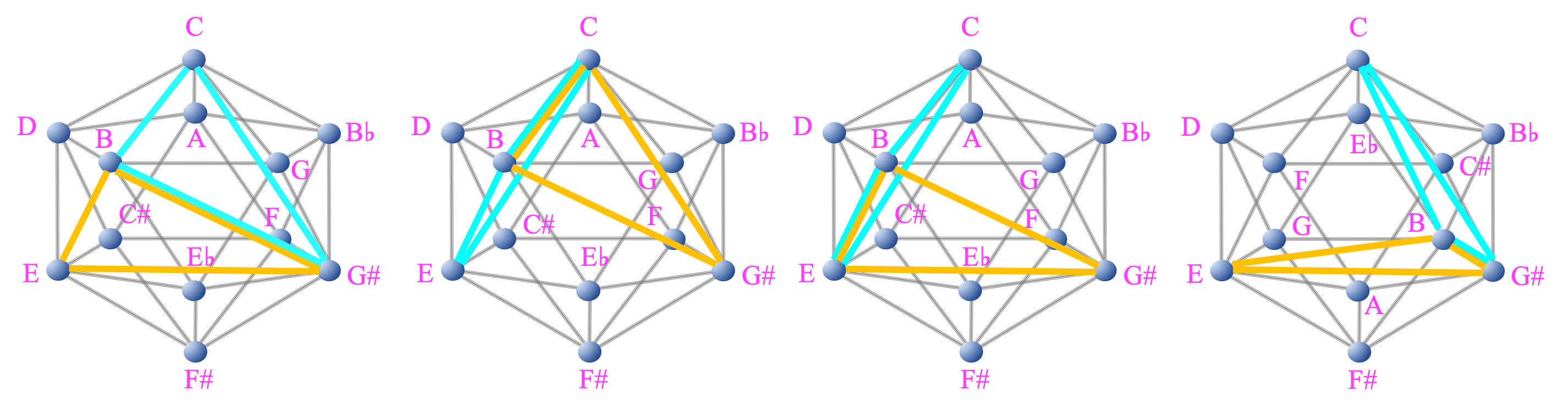}}}\hspace{5pt}
\caption{The golden analysis of $C$, $E$, $G\sharp$, $B$. There exist three golden decompostions of $C$, $E$, $G\sharp$, $B$, for the type ${\rm 1^*}$: $C$, $E$, $G\sharp$, $B$ = $C$, $G\sharp$, $B$ (gt) + $E$, $G\sharp$, $B$ (gt), $C$, $E$, $G\sharp$, $B$ = $C$, $E$, $B$ (gg) + $C$, $G\sharp$, $B$ (gt), $C$, $E$, $G\sharp$, $B$ = $C$, $E$, $B$ (gg) + $E$, $G\sharp$, $B$ (gt). There exists one golden decompostion of $C$, $E$, $G\sharp$, $B$, for the type ${\rm 2^*}$: $C$, $E$, $G\sharp$, $B$ = $C$, $G\sharp$, $B$ (gt) + $E$, $G\sharp$, $B$ (gt).}
\label{CEGsB}
\end{figure}

In this subsection, we show results of the golden analysis for the tertian seventh chords: the major seventh chord ($C$, $E$, $G$, $B$, Fig.\,\ref{CEGB}), the minor seventh chord ($C$, $E\flat$, $G$, $B\flat$, Fig.\,\ref{CEfGBf}), the dominant seventh chord ($C$, $E$, $G$, $B\flat$, Fig.\,\ref{CEGBf}), the diminished seventh chord ($C$, $E\flat$, $G\flat$, $A$, Fig.\,\ref{CEfGfA}), the half-diminished seventh chord ($C$, $E\flat$, $G\flat$, $B\flat$, Fig.\,\ref{CEfGfBf}), the minor major seventh chord ($C$, $E\flat$, $G$, $B$, Fig.\,\ref{CEfGB}), and the augmented major seventh chord ($C$, $E$, $G\sharp$, $B$, Fig.\,\ref{CEGsB}). In captions of the figures, we use an algebra: $X_1, \ X_2, \ \cdots, X_n = X_{n_1}, \cdots, X_{n_k} + X_{n_{k+1}}, \cdots, X_{n_m}$ if $X_1, \ X_2, \ \cdots, X_n$ has golden decompositions, $ X_{n_1}, \cdots, X_{n_k}$ and $X_{n_{k+1}}, \cdots, X_{n_m}$, and use the following abbreviations: gt (golden triangle) and gg (golden gnomon).

One can see the dominant seventh chord ($C$, $E$, $G$, $B\flat$) is the only tertian seventh chord that is golden singular in the type ${\rm 1^*}$ and the half-diminished seventh chord ($C$, $E\flat$, $G\flat$, $B\flat$) is the only tertian seventh chord that is golden singular in the type ${\rm 2^*}$. Therefore, from the viewpoint of golden singular on the exceptional musical icosahedra, the dominant seventh chord is the dual seventh chord of the half-diminished seventh chord.
\\
\\
\indent
[{\bf Golden Singular Duality}]

In the type ${\rm 1^*}$ and type ${\rm 4^*}$, the dominant seventh chord is the only tertian seventh chord that is golden singular.

In the type ${\rm 2^*}$ and type ${\rm 3^*}$, the half-diminished seventh chord is the only tertian seventh chord that is golden singular.
\\
\\
\indent
Note that a figure corresponding to the dominant seventh chord in the type ${\rm 1^*}$ is the same as a figure corresponding to the half-diminished seventh chord in the type ${\rm 2^*}$. In fact, $C$, $F\sharp$, $A$, $B\flat$ is obtained by a symmetry operation to a figure corresponding to $C$, $E$, $G$, $B\flat$ (the dominant seventh chord), and a transformation of $C$, $F\sharp$, $A$, $B\flat$ from $A$ to $E\flat$ leads to $C$, $E\flat$, $F\sharp$, $B\flat$ (the half-diminished seventh chord). Note that the type ${\rm 2^*}$ is obtained by applying $C\leftrightarrow F\sharp$, $C\sharp\leftrightarrow G$, $D\leftrightarrow G\sharp$, $E\flat\leftrightarrow A$, $E\sharp\leftrightarrow B\flat$, $F\leftrightarrow B$ to the type ${\rm 1^*}$.

\begin{figure}[H]
\centering
{%
\resizebox*{16cm}{!}{\includegraphics{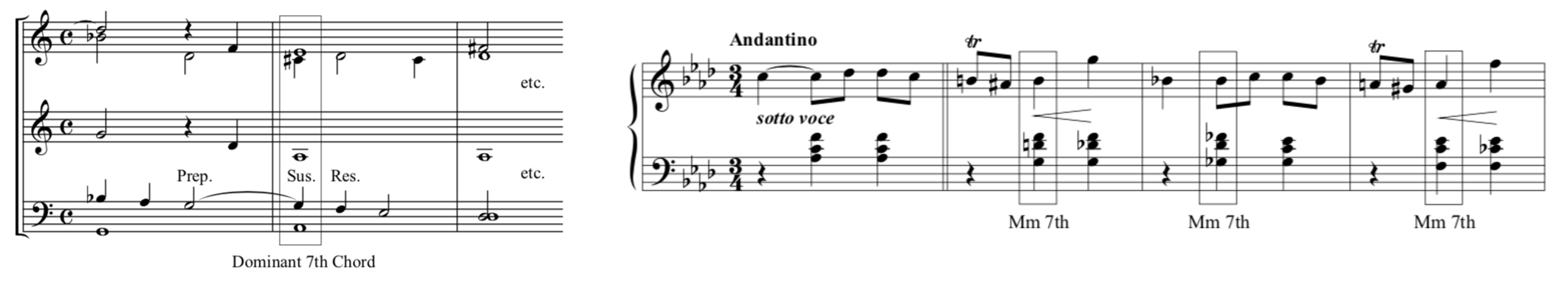}}}\hspace{5pt}
\caption{Monteverdi (1567-1643): ``Lasciatemi morire" (``Oh, Let Me Die") from Lamento d’Arianna, mm. 6–8. and Chopin (1810-1849): Mazurka in F Minor, op. posth. 68, no. 4, mm. 1–4. (Benward, Saker, 2009). Mm 7th means the dominant seventh chord.}
\label{score1}
\end{figure}

\begin{figure}[H]
\centering
{%
\resizebox*{8cm}{!}{\includegraphics{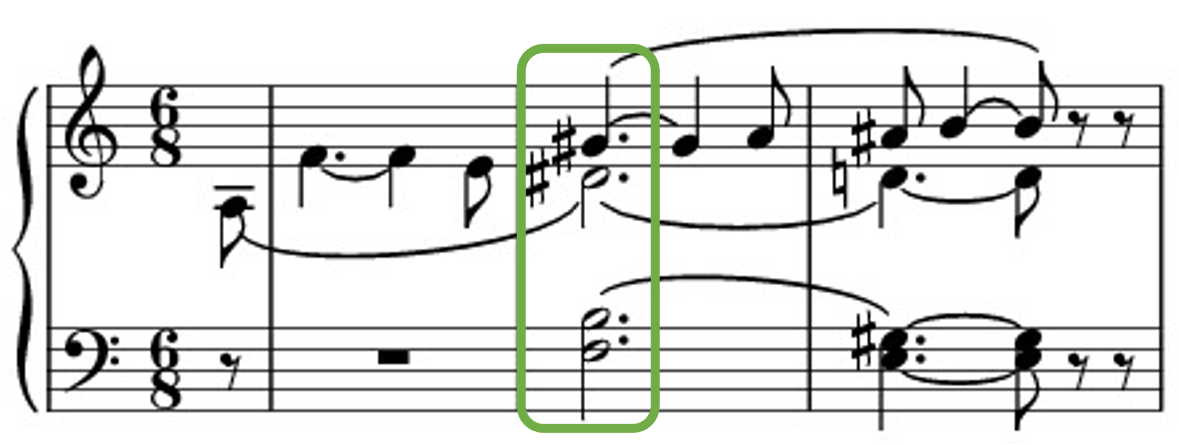}}}\hspace{5pt}
\caption{The first part of the piano reduction of \emph{Tristan und Isolde} includes a Tristan chord, $F$, $B$, $D\sharp$, $G\sharp$.}
\label{score2}
\end{figure}

These two seventh chords can be regarded as the most important seventh chords among seventh chords. The importance of the dominant seventh chord was shown in detail by Benward and Saker (Benward, Saker, 2009). They wrote ``Early baroque period composers, such as Monteverdi and Scheidt, introduced the V7 [dominant seventh] chord...Later in the baroque period, V7 chords were more plentiful and became an integral part of the musical language. The dominant seventh chord was in constant use throughout the classical period...In the romantic period, dominant seventh chords were plentiful, but freer voice-leading treatment gradually developed." They also showed how the dominant seventh chord has been used in famous musical pieces (Fig.\,\ref{score1}). Also, the half-diminisehd chord played an important role in Richard Wagner's \emph{Tristan und Isolde}, and called the Tristan chord in the context of Tristan und Isolde. Martin pointed out in his analysis of the Tristan chord (Martin, 2008) that the Tristan chord has attracted music theorist's attention, more than a hundred years, and quoted Robert Wason's notation (Wason, 1982), the Tristan chord has almost seemed to serve as a touchstone against which any theory of harmony must prove itself. The first part of the piano reduction of \emph{Tristan und Isolde}, and the Tristan chord is shown in Fig.\,\ref{score2}. It is interesting that these two historically important seventh chords are special in terms of the golden singularity in our theory.

We also remarked that the golden analysis reveals a similarity between the minor major seventh chord and the augmented major seventh chord. While the minor major seventh chord has three golden decompositions for the type ${\rm 1^*}$ and one golden decomposition the type ${\rm 2^*}$, the augmented major seventh chord has one golden decomposition for the type ${\rm 1^*}$ has three golden decompositions for the type ${\rm 2^*}$. The similarity between these two seventh chords leads to a tetrad analogue of the duality between the major triad and the minor triad. By applying a proper transposition and interchange, we have $C$, $G\sharp$, $E$, $C\sharp$ from $C$, $E\flat$, $G$, $B$, and $B\flat$, $G$, $E\flat$, $B$ from $C$, $E$, $G\sharp$, $B$. Then, one can see the similarity between $C$, $G\sharp$, $E$, $C\sharp$ and $C$, $E$, $G\sharp$, $B$, and $B\flat$, $G$, $E\flat$, $B$ and $C$, $G\sharp$, $E$, $C\sharp$. The $C$, $G\sharp$, $E$, $C\sharp$ is obtained by interchanging the second tone and the third tone, and raising the final tone by two semitones in $C$, $E$, $G\sharp$, $B$ (augmented major seventh chord). The $B\flat$, $G$, $E\flat$, $B$ is obtained by interchanging the second tone and the third tone, and raising the first tone by two semitones in  $C$, $E\flat$, $G$, $B$ (minor major seventh chord). The remarkable point is that for a golden-base harmony of $C$, $G\sharp$, $E$, $C\sharp$ ($B\flat$, $G$, $E\flat$, $B$) that is represented by a golden triangle/gnomon, the naturally corresponding golden-base harmony\footnote{For a golden-base harmony $X_{n_1}$, $X_{n_2}$, $X_{n_3}$ for a harmony $X_1$, $X_2$, $X_3$, $X_4$, the naturally corresponding golden-base harmony is $Y_{n_1}$, $Y_{n_2}$, $Y_{n_3}$ for a harmony $Y_1$, $Y_2$, $Y_3$, $Y_4$ ($n_1$, $n_2$, $n_3$, $n_4$ are natural numbers less than 5).} of $C$, $E$, $G\sharp$, $B$ ($C$, $E\flat$, $G$, $B$) is represented by a golden gnomon/triangle. We also remark that the major triad $C$, $E$, $G$ (minor triad $C$, $E\flat$, $G$) is given by raising $E\flat$ (lowering $E$) by one semitone in the minor triad (major triad). Also, for a golden-base harmony of $C$, $E$, $G$ that is represented by a golden triangle/gnomon, the naturally corresponding golden-base harmony of $C$, $E\flat$, $G$ is represented by a golden gnomon/triangle\footnote{Note that a golden-base harmony of $C$, $E$, $G$ ($C$, $E\flat$, $G$) is $C$, $E$, $G$ ($C$, $E\flat$, $G$) itself.}.

Therefore, from the viewpoint of the golden analysis, a relation between the major triad and the minor triad can be generalized. A harmony $A$ is a generalized major-minor dual of a harmony $B$ if and only if (i) $A$ is obtained by lowering/raising of one tone constructing $B$ by some semitones and interchanging some tones constructing $B$, (ii) for a golden-base harmony of $A$ that is represented by a golden triangle/gnomon, the naturally corresponding golden-base harmony of $B$ is represented by a golden gnomon/triangle. Obviously, the $X$ major triad is a generalized major-minor dual of the $X$ minor triad. Then, we have the following theorem.
\\
\\
\indent
[{\bf Generalized Major-Minor Duality in Seventh Chords}]

The generalized major-minor dual of $C$, $E\flat$, $G$, $B$ (minor major seventh chord) is $B\flat$, $G$, $E\flat$, $B$ (a rearrangement of transposition of the augmented major seventh chord).

\begin{figure}[H]
\centering
{%
\resizebox*{5cm}{!}{\includegraphics{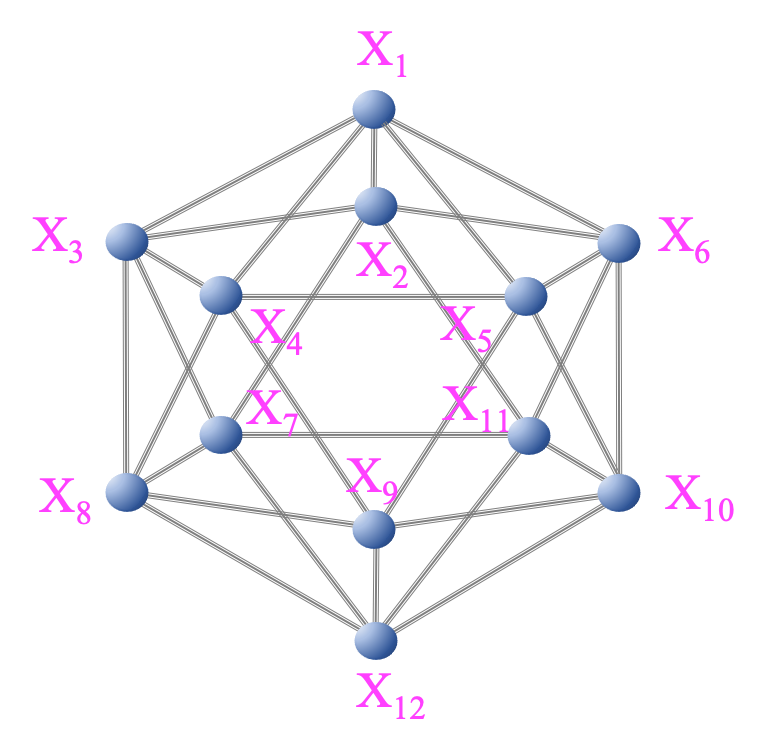}}}\hspace{5pt}
\caption{A musical icosahedron with $X_1$, $X_2$, $X_3$, $X_4$, $X_5$, $X_6$, $X_7$, $X_8$, $X_9$, $X_{10}$, $X_{11}$, $X_{12}$.}
\label{five_harmony}
\end{figure}

The generalized major-minor dual of $C$, $E$, $G\sharp$, $B$ (augmented major seventh chord) and $C$, $G\sharp$, $E$, $C\sharp$ (a rearrangement of transposition of the minor major seventh chord).
\\
\\
\indent
The golden analysis is also applied to harmonies constructed by five or more tones. One can prove that a harmony constructed by five or more tones is not golden singular. Because a harmony constructed by six or more tones can be decomposed into some harmonies constructed by five tones, it is enough to deal with only a harmony constructed by five tones in order to prove the above proposition. The number of combinations of choosing 5 tones from 12 tones is 792. However, by considering the symmetry of the regular icosahedron, the number of combinations we should deal with can be reduced. In order to explain the proof, we use a musical icosahedron shown in Fig.\,\ref{five_harmony}. First, we deal with harmonies including $X_1$ and $X_{12}$. In this case, it is enough to deal with the following patterns: $X_2$, $X_3$, $X_4$ and $X_2$, $X_3$, $X_5$ and $X_2$, $X_3$, $X_8$ and $X_2$, $X_3$, $X_9$ and $X_2$, $X_3$, $X_{10}$ and $X_2$, $X_4$, $X_8$ and $X_2$, $X_4$, $X_8$ and $X_2$, $X_4$, $X_9$ and $X_2$, $X_4$, $X_{10}$. Then, it is obvious they are not golden singular. Second, we deal with harmonies including $X_1$ and not including $X_{12}$. In this case, it is enough to deal with the following patterns: $X_2$, $X_3$, $X_4$, $X_5$ and $X_2$, $X_3$, $X_4$, $X_8$ and $X_2$, $X_3$, $X_4$, $X_9$ and $X_2$, $X_3$, $X_4$, $X_{10}$ and $X_2$, $X_3$, $X_5$, $X_8$ and $X_2$, $X_3$, $X_5$, $X_9$ and $X_2$, $X_3$, $X_5$, $X_{10}$ and $X_2$, $X_3$, $X_7$, $X_8$ and $X_2$, $X_3$, $X_8$, $X_9$ and $X_2$, $X_3$, $X_9$, $X_{10}$ and $X_2$, $X_3$, $X_7$, $X_9$ and $X_2$, $X_3$, $X_8$, $X_{10}$ and $X_2$, $X_3$, $X_8$, $X_{11}$ and $X_2$, $X_7$, $X_8$, $X_9$ and $X_2$, $X_8$, $X_9$, $X_{10}$ and $X_2$, $X_7$, $X_8$, $X_{11}$ and $X_2$, $X_7$, $X_8$, $X_{10}$ and $X_2$, $X_8$, $X_9$, $X_{11}$ and $X_2$, $X_7$, $X_9$, $X_{11}$. Then, it is obvious they are not golden singular.

\begin{figure}[H]
\centering
{%
\resizebox*{16cm}{!}{\includegraphics{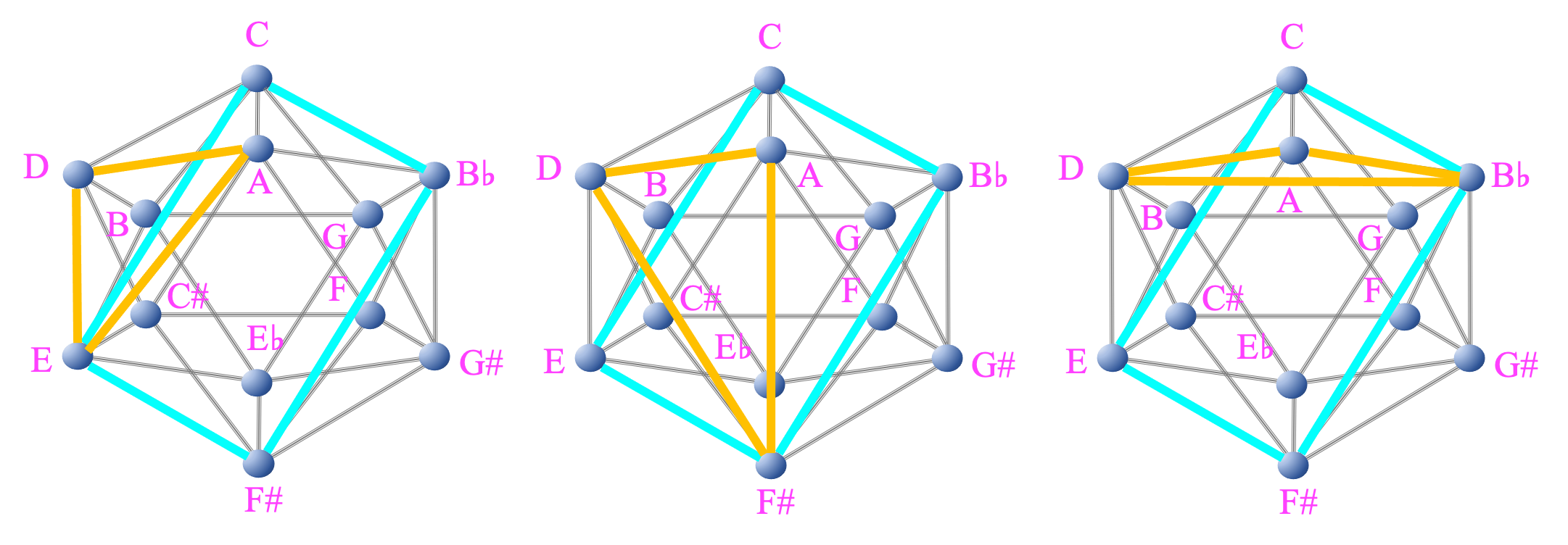}}}\hspace{5pt}
\caption{The golden analysis of the mystic chord by the type ${\rm 1^*}$.}
\label{mystic1}
\end{figure}

\begin{figure}[H]
\centering
{%
\resizebox*{16cm}{!}{\includegraphics{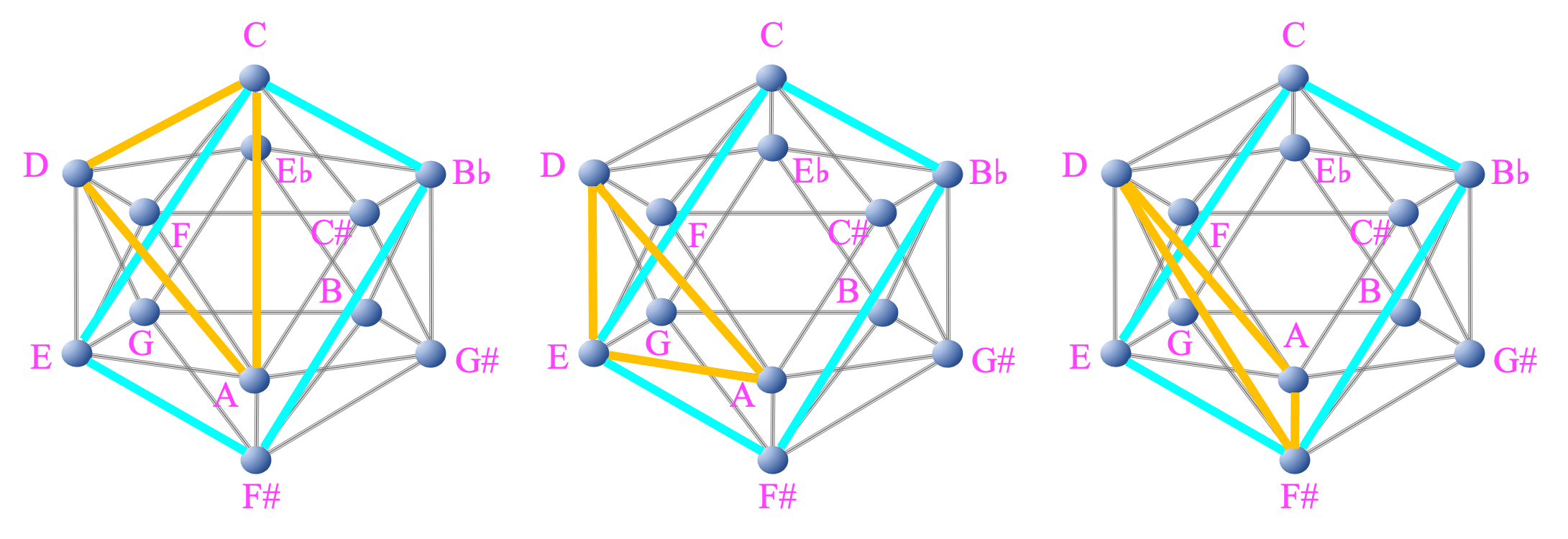}}}\hspace{5pt}
\caption{The golden analysis of the mystic chord by the type ${\rm 2^*}$.}
\label{mystic2}
\end{figure}

For example, we show the results of the golden analysis of the mystic chord that was often used by Scriabin (Fig.\,\ref{mystic1} and Fig.\,\ref{mystic2}). The mystic chord can be represented by a golden triangle/gnomon and a golden rectangle. Both the type ${\rm 1^*}$ and the type ${\rm 2^*}$ have three golden decompositions.

\newpage
\section{Golden Analysis of BWV 846}
In this section, we show the results of the golden analysis of the BWV 846 composed by Johann Sebastian Bach. Because each measure of this piece is constructed by a simple harmony, the golden analysis can be applied. Note that if a harmony $C$, $E$, $G\sharp$, or $C$, $F\sharp$, $E$, etc. is included in a given musical piece, one cannot use the golden analysis for that piece. Because the BWV 846 includes the dominant seventh chord, we should use the type ${\rm 2^*}$ and/or the type ${\rm 3^*}$. In Fig.\,\ref{Bach1} - Fig.\,\ref{Bach28}, we show how the type ${\rm 2^*}$ and the type ${\rm 3^*}$ dually represent the BWV 846 by the golden ratio on the regular icosahedron. Note that if a harmony in a measure of the BWV 846 has several golden decompositions, we choose one of those golden decompositions to minimize the total combinations of the golden figures representing the BWV 846. The BWV 846 can be represented by only 7 combinations of the golden figures on the type ${\rm 2^*}$ or the type ${\rm 3^*}$: a golden triangle, a golden gnomon, a golden rectangle, two golden triangles, two golden gnomons, a golden rectangle and a golden triangle, a golden rectangle and a golden gnomon.

The measure 1, 4, 15, 19, 25, 29, 35 of BWV 846 are characterized by $C$, $E$, $G$ corresponding to a golden triangle on the type ${\rm 2^*}$ and a golden gnomon on the type ${\rm 3^*}$ (Fig.\,\ref{Bach1}). The measure  2, 17, 33 of BWV 846 are characterized by $D$, $F$, $A$, $C$ corresponding to two golden triangles ($D$, $F$, $A$ and $F$, $A$, $C$) on the type ${\rm 2^*}$ and two golden gnomons ($D$, $F$, $A$ and $F$, $A$, $C$) on the type ${\rm 3^*}$ (Fig.\,\ref{Bach2}). The measure 3, 7, 18, 24, 27, 31 of BWV 846 are characterized by $G$, $B$, $D$, $F$ corresponding to two golden gnomons ($G$, $B$, $D$ and $G$, $D$, $F$) on the type ${\rm 2^*}$ and two golden triangles ($G$, $B$, $D$ and $G$, $D$, $F$) on the type ${\rm 3^*}$ (Fig.\,\ref{Bach3}). The measure 5 of BWV 846 is characterized by $A$, $C$, $E$ corresponding to a golden triangle on the type ${\rm 2^*}$ and a golden gnomon on the type ${\rm 3^*}$. The measure 6, 10 of BWV 846 are characterized by $D$, $F\sharp$, $A$, $C$ corresponding to two golden triangles ($D$, $F\sharp$, $A$ and $D$, $A$, $C$) on the type ${\rm 2^*}$ and two golden gnomons ($D$, $F\sharp$, $A$ and $D$, $A$, $C$) on the type ${\rm 3^*}$. The measure 8 of BWV 846 is characterized by $C$, $E$, $G$, $B$ corresponding to two golden triangles ($C$, $E$, $G$ and $C$, $G$, $B$) on the type ${\rm 2^*}$ and two golden gnomons ($C$, $E$, $G$ and $C$, $G$, $B$) on the type ${\rm 3^*}$. The measure 9 of BWV 846 is characterized by $C$, $E$, $G$, $A$ corresponding to two golden triangles ($C$, $E$, $G$ and $A$, $C$, $E$) on the type ${\rm 2^*}$ and two golden gnomons ($C$, $E$, $G$ and $A$, $C$, $E$) on the type ${\rm 3^*}$. The measure 11 of BWV 846 is characterized by $G$, $B$, $D$ corresponding to a golden gnomon on the type ${\rm 2^*}$ and a golden triangle on the type ${\rm 3^*}$. The measure 12 of BWV 846 is characterized by $C\sharp$, $E$, $G$, $B\flat$ corresponding to a golden rectangle on the type ${\rm 2^*}$ and type ${\rm 3^*}$. The measure 13 of BWV 846 is characterized by $D$, $F$, $A$ corresponding to a golden gnomon on the type ${\rm 2^*}$ and a golden triangle on the type ${\rm 3^*}$. The measure 16, 21 of BWV 846 are characterized by $F$, $A$, $C$, $E$ corresponding to two golden gnomons ($F$, $A$, $C$ and $F$, $C$, $E$) on the type ${\rm 2^*}$ and two golden triangles ($F$, $A$, $C$ and $F$, $C$, $E$) on the type ${\rm 3^*}$. The measure 20, 32 of BWV 846 are characterized by $C$, $E$, $G$, $B\flat$ corresponding to two golden triangles ($C$, $E$, $G$ and $C$, $E$, $B\flat$) on the type ${\rm 2^*}$ and two golden gnomons ($C$, $E$, $G$ and $C$, $E$, $B\flat$) on the type ${\rm 3^*}$. The measure 22 of BWV 846 is characterized by $C$, $E\flat$, $F\sharp$, $A$ corresponding to a golden rectangle on the type ${\rm 2^*}$ and type ${\rm 3^*}$. The measure 23 of BWV 846 is characterized by $D$, $F$, $A\flat$, $B$, $C$ corresponding to a golden rectangle and a golden triangle ($D$, $F$, $A\flat$, $B$ and $F$, $A\flat$, $C$) on the type ${\rm 2^*}$ and a golden rectangle and a golden gnomon ($D$, $F$, $A\flat$, $B$ and $F$, $A\flat$, $C$) on the type ${\rm 3^*}$. The measure 26 of BWV 846 is characterized by $G$, $D$, $C$, $F$ corresponding to two golden gnomons ($C$, $F$, $G$ and $G$, $D$, $F$) on the type ${\rm 2^*}$ and two golden triangles ($C$, $F$, $G$ and $G$, $D$, $F$) on the type ${\rm 3^*}$. The measure 28 of BWV 846 is characterized by $C$, $E\flat$, $F\sharp$, $G$, $A$ corresponding to a golden rectangle and a golden gnomon ($C$, $E\flat$, $F\sharp$, $A$ and $C$, $E\flat$, $G$) on the type ${\rm 2^*}$ and a golden rectangle and a golden triangle ($C$, $E\flat$, $F\sharp$, $A$ and $C$, $E\flat$, $G$) on the type ${\rm 3^*}$. The measure 34 of BWV 846 is characterized by $C$, $D$, $E$, $F$ $G$, $B$ corresponding to two golden gnomons ($G$, $B$, $D$ and $C$, $E$, $F$) on the type ${\rm 2^*}$ and two golden triangles ($G$, $B$, $D$ and $C$, $E$, $F$) on the type ${\rm 3^*}$.

\begin{figure}[H]
\centering
{%
\resizebox*{10cm}{!}{\includegraphics{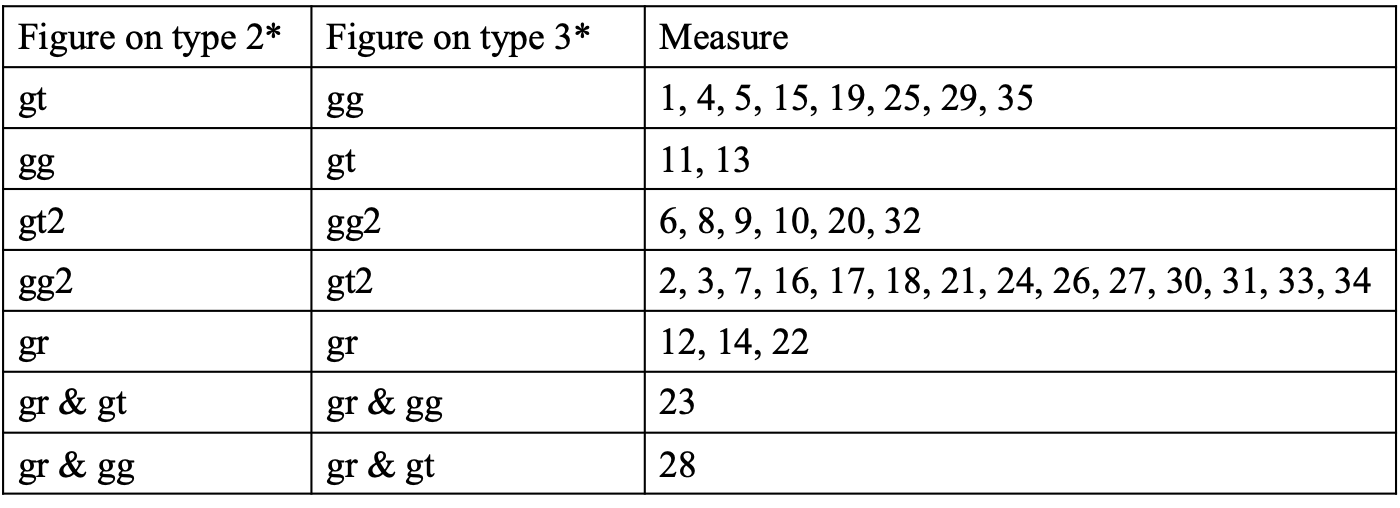}}}\hspace{5pt}
\caption{Summary of the golden analysis of the BWV 846. Seven combinations of the golden figures (a golden triangle, a golden gnomon, a golden rectangle, two golden triangles, two golden gnomons, a golden rectangle and a golden triangle, a golden rectangle and a golden gnomon) characterize all the measures. The term ``gt/gg" means a golden triangle/gnomon, the term ``gt2/gg2" means two golden triangles/gnomons, the term ``gr" means a golden rectangle, and the term ``gr \& gt/gg" means a golden rectangle and golden triangle/gnomon.}
\label{summary_golden}
\end{figure}

\begin{figure}[H]
\centering
{%
\resizebox*{16cm}{!}{\includegraphics{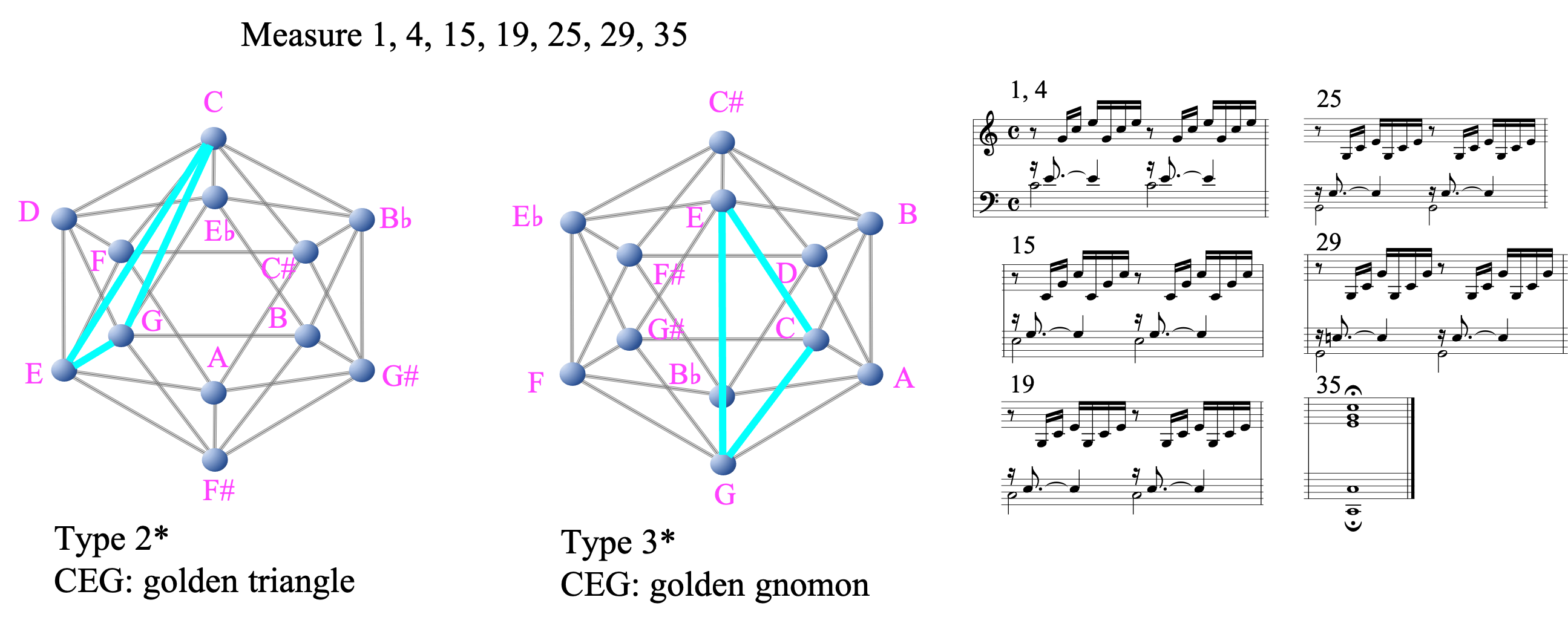}}}\hspace{5pt}
\caption{The golden analysis of the measure 1, 4, 15, 19, 25, 29, 35 of BWV 846 on the type ${\rm 2^*}$ and type ${\rm 3^*}$. $C$, $E$, $G$ is represented by a golden triangle on the type ${\rm 2^*}$ and a golden gnomon on the type ${\rm 3^*}$.}
\label{Bach1}
\end{figure}

\begin{figure}[H]
\centering
{%
\resizebox*{16cm}{!}{\includegraphics{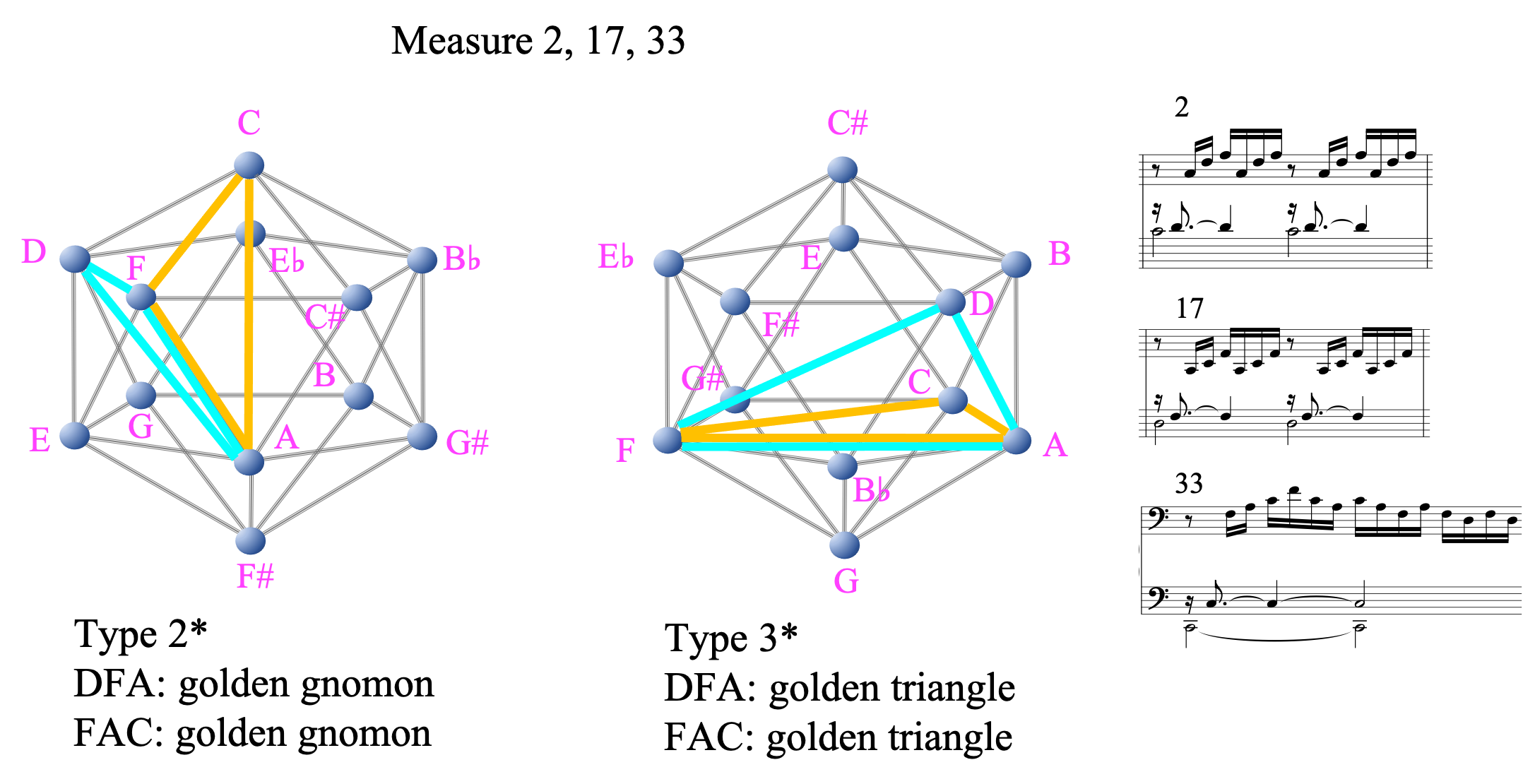}}}\hspace{5pt}
\caption{The golden analysis of the measure 2, 17, 33 of BWV 846 on the type ${\rm 2^*}$ and type ${\rm 3^*}$. $D$, $F$, $A$, $C$ is represented by two golden triangles ($D$, $F$, $A$ and $F$, $A$, $C$) on the type ${\rm 2^*}$ and two golden gnomons ($D$, $F$, $A$ and $F$, $A$, $C$) on the type ${\rm 3^*}$. }
\label{Bach2}
\end{figure}

\begin{figure}[H]
\centering
{%
\resizebox*{17cm}{!}{\includegraphics{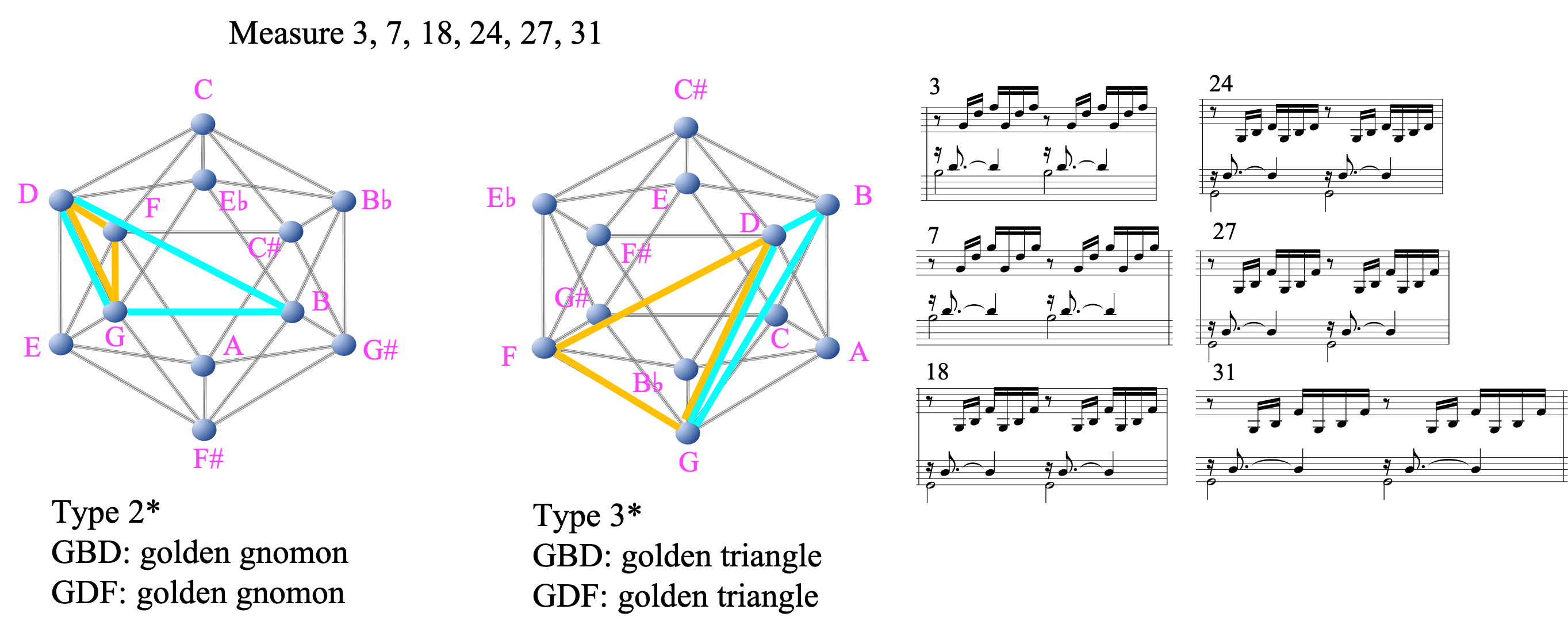}}}\hspace{5pt}
\caption{The golden analysis of the measure 3, 7, 18, 24, 27, 31 of BWV 846 on the type ${\rm 2^*}$ and type ${\rm 3^*}$. $G$, $B$, $D$, $F$ is represented by two golden gnomons ($G$, $B$, $D$ and $G$, $D$, $F$) on the type ${\rm 2^*}$ and two golden triangles ($G$, $B$, $D$ and $G$, $D$, $F$) on the type ${\rm 3^*}$. }
\label{Bach3}
\end{figure}

Figure\,\ref{summary_golden} shows a summary of the golden analysis of the BWV 846. It is found that a combination of two golden gnomons (triangles) on the type ${\rm 2^*}$ (type ${\rm 3^*}$) is the most frequent combination among the combinations of the golden figures representing the measures of the BWV 846.

\begin{figure}[H]
\centering
{%
\resizebox*{16cm}{!}{\includegraphics{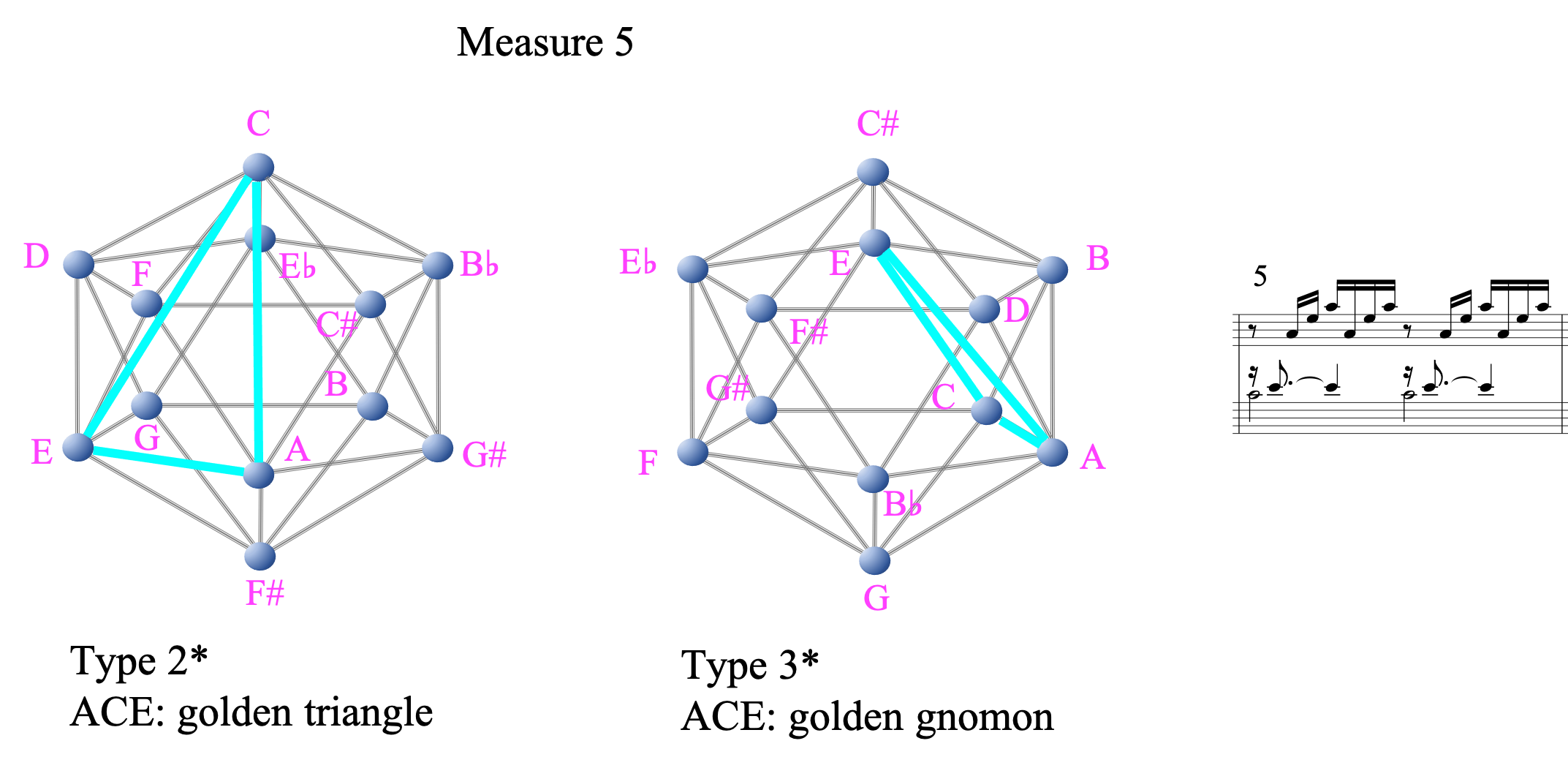}}}\hspace{5pt}
\caption{The golden analysis of the measure 5 of BWV 846 on the type ${\rm 2^*}$ and type ${\rm 3^*}$. $A$, $C$, $E$ is represented by a golden triangle on the type ${\rm 2^*}$ and a golden gnomon on the type ${\rm 3^*}$. }
\label{Bach5}
\end{figure}

\begin{figure}[H]
\centering
{%
\resizebox*{16cm}{!}{\includegraphics{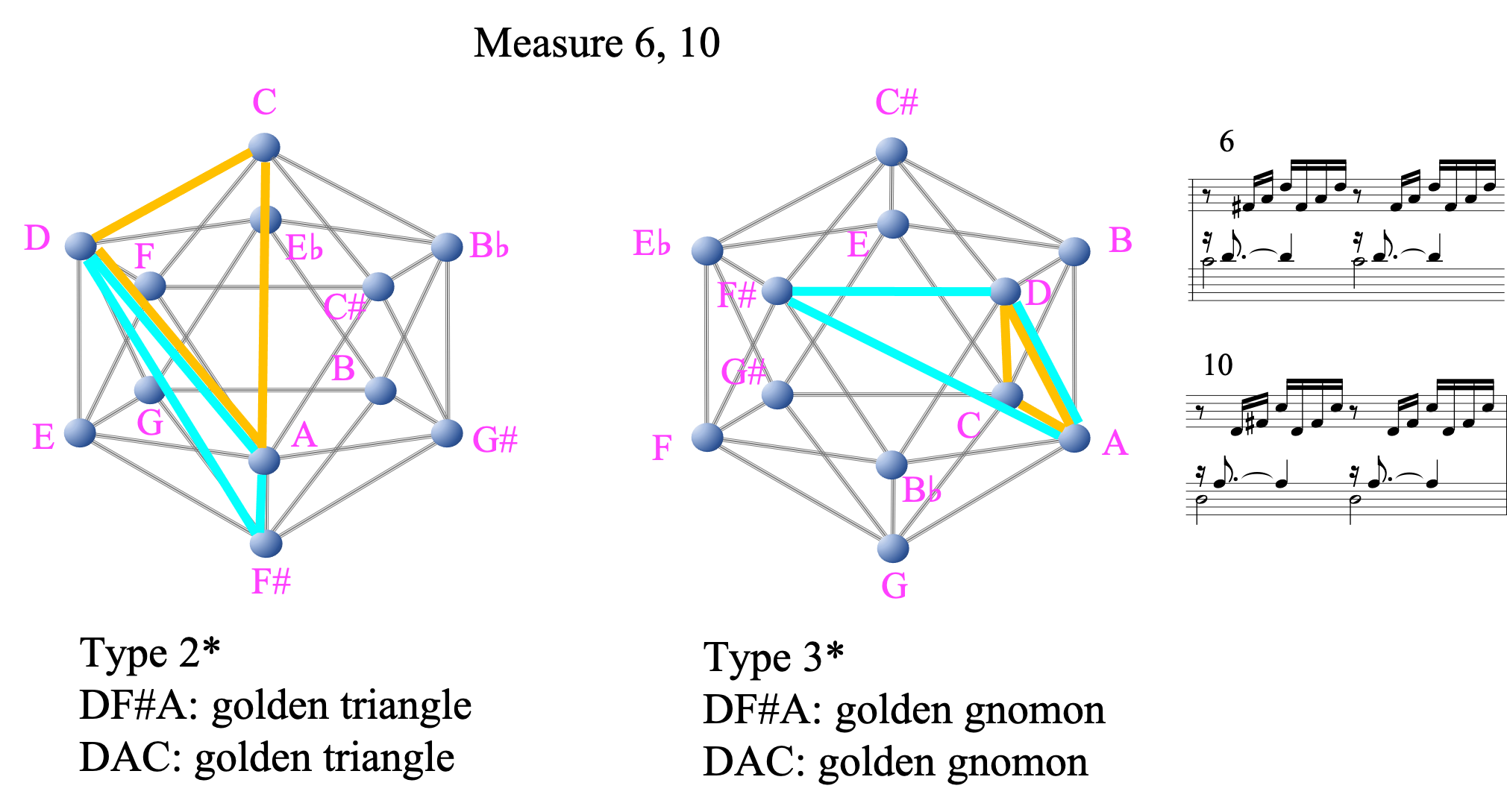}}}\hspace{5pt}
\caption{The golden analysis of the measure 6, 10 of BWV 846 on the type ${\rm 2^*}$ and type ${\rm 3^*}$. $D$, $F\sharp$, $A$, $C$ is represented by two golden triangles ($D$, $F\sharp$, $A$ and $D$, $A$, $C$) on the type ${\rm 2^*}$ and two golden gnomons ($D$, $F\sharp$, $A$ and $D$, $A$, $C$) on the type ${\rm 3^*}$. }
\label{Bach6}
\end{figure}

\begin{figure}[H]
\centering
{%
\resizebox*{16cm}{!}{\includegraphics{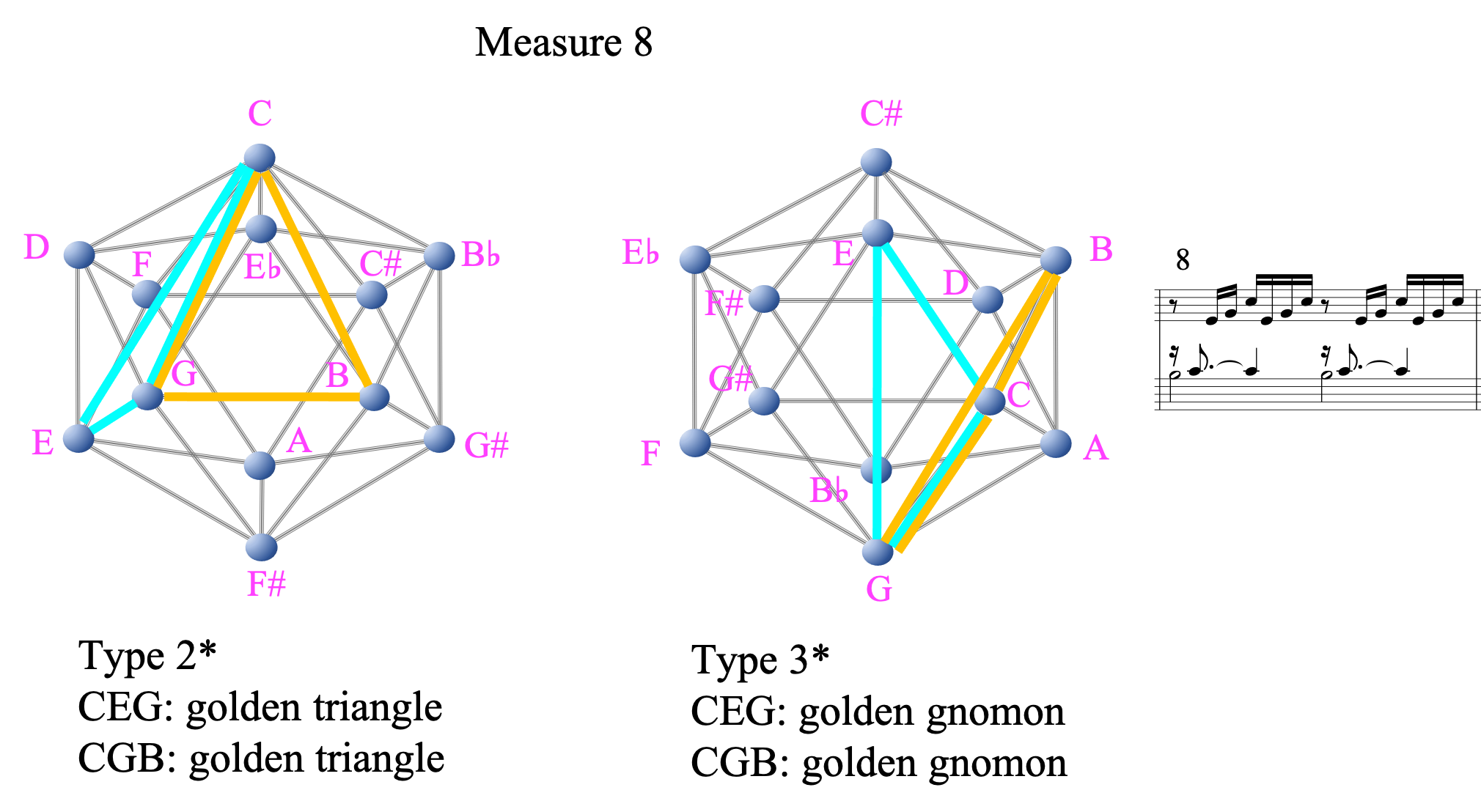}}}\hspace{5pt}
\caption{The golden analysis of the measure 8 of BWV 846 on the type ${\rm 2^*}$ and type ${\rm 3^*}$. $C$, $E$, $G$, $B$ is represented by two golden triangles ($C$, $E$, $G$ and $C$, $G$, $B$) on the type ${\rm 2^*}$ and two golden gnomons ($C$, $E$, $G$ and $C$, $G$, $B$) on the type ${\rm 3^*}$. }
\label{Bach8}
\end{figure}

\begin{figure}[H]
\centering
{%
\resizebox*{16cm}{!}{\includegraphics{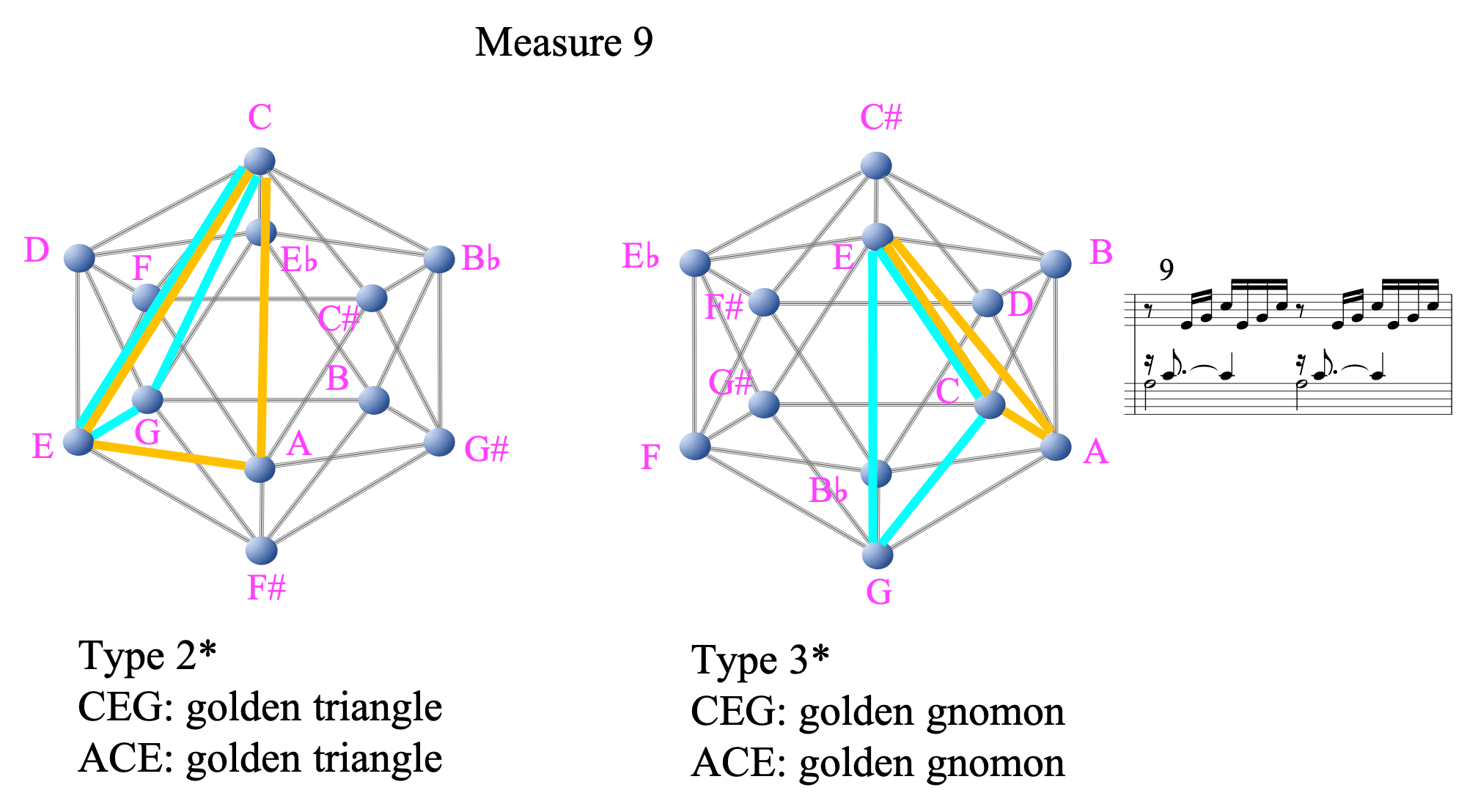}}}\hspace{5pt}
\caption{The golden analysis of the measure 9 of BWV 846 on the type ${\rm 2^*}$ and type ${\rm 3^*}$. $C$, $E$, $G$, $A$ is represented by two golden triangles ($C$, $E$, $G$ and $A$, $C$, $E$) on the type ${\rm 2^*}$ and two golden gnomons ($C$, $E$, $G$ and $A$, $C$, $E$) on the type ${\rm 3^*}$. }
\label{Bach9}
\end{figure}

\begin{figure}[H]
\centering
{%
\resizebox*{16cm}{!}{\includegraphics{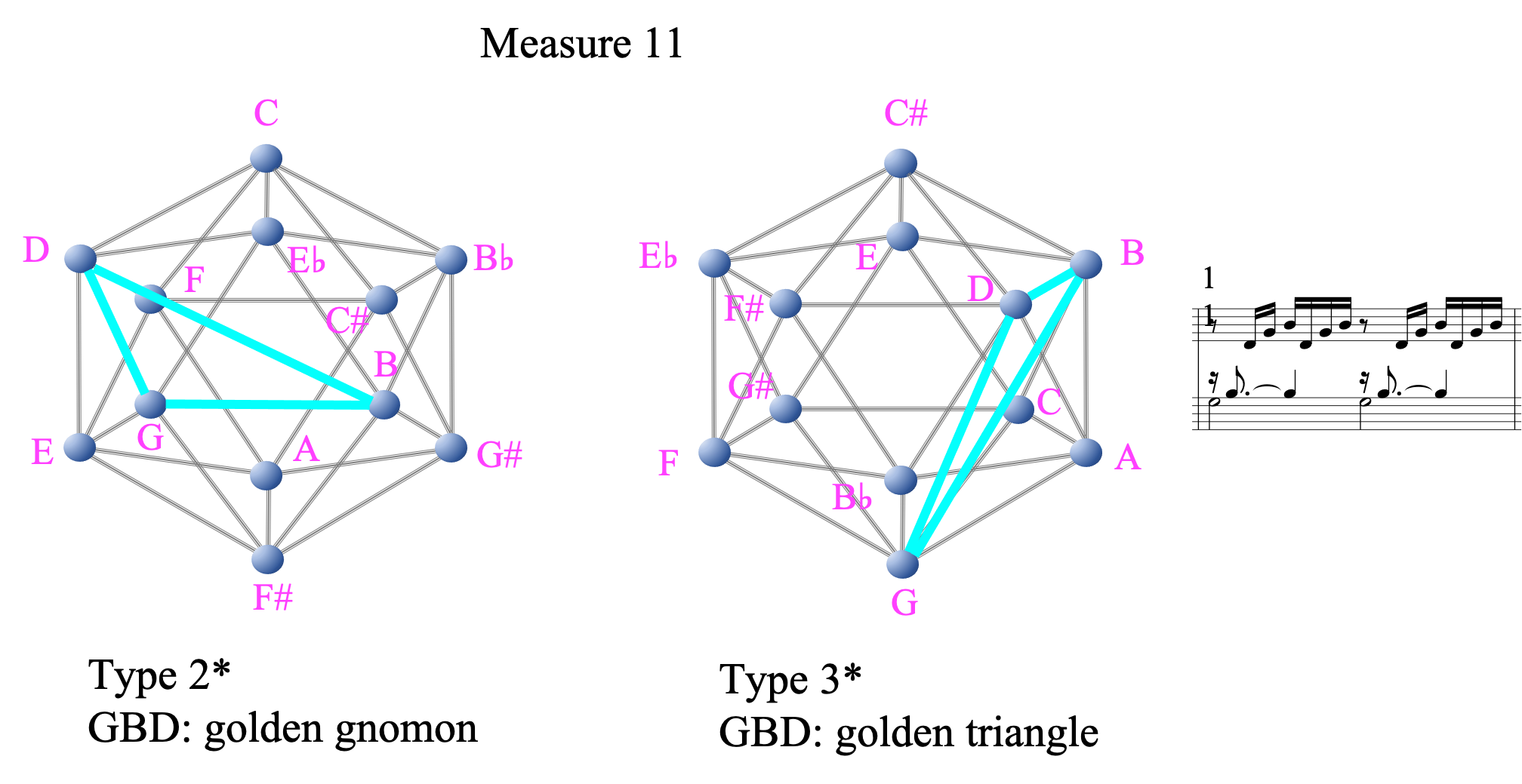}}}\hspace{5pt}
\caption{The golden analysis of the measure 11 of BWV 846 on the type ${\rm 2^*}$ and type ${\rm 3^*}$. $G$, $B$, $D$ is represented by a golden gnomon on the type ${\rm 2^*}$ and a golden triangle on the type ${\rm 3^*}$. }
\label{Bach11}
\end{figure}

\begin{figure}[H]
\centering
{%
\resizebox*{16cm}{!}{\includegraphics{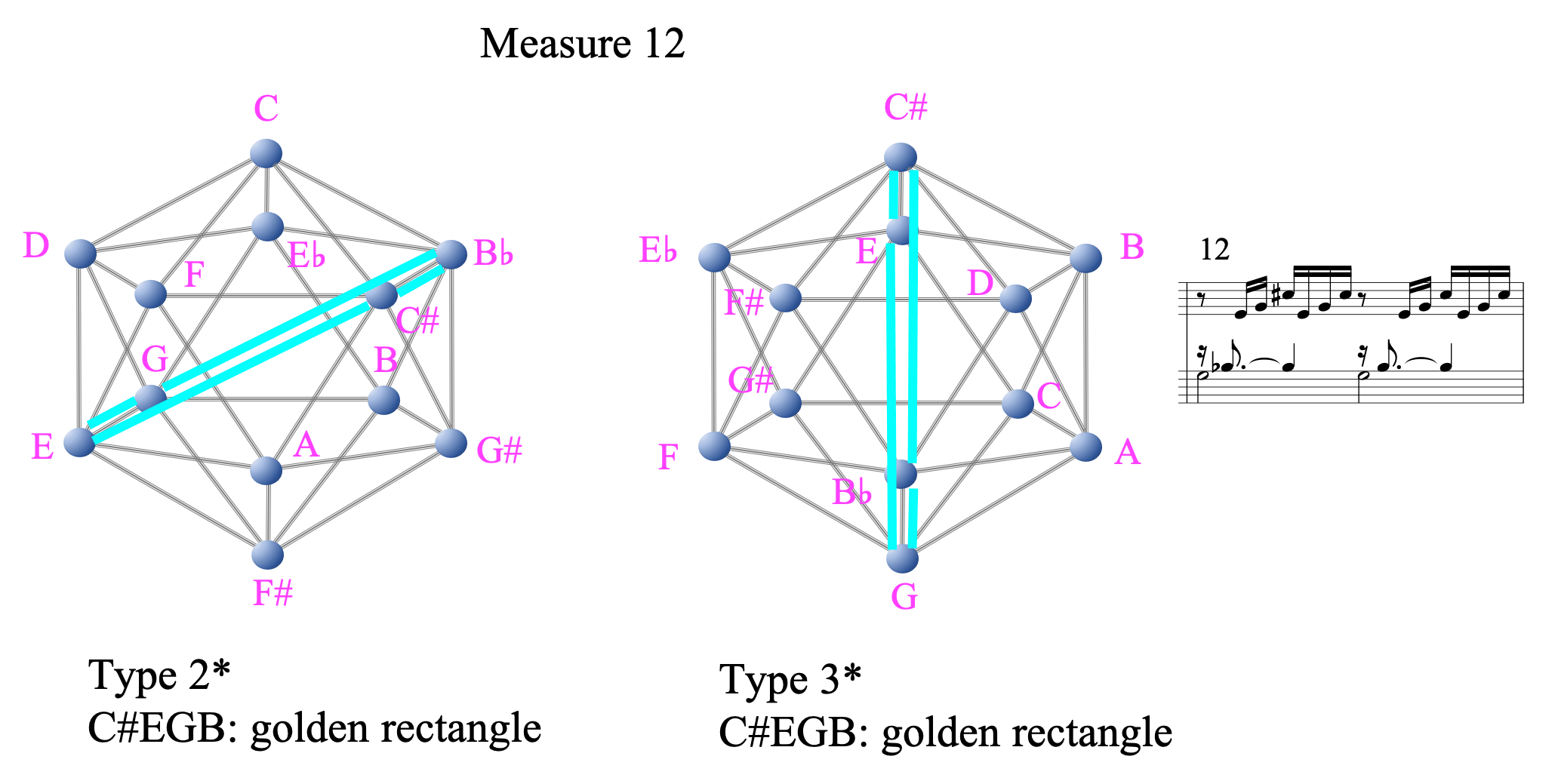}}}\hspace{5pt}
\caption{The golden analysis of the measure 12 of BWV 846 on the type ${\rm 2^*}$ and type ${\rm 3^*}$. $C\sharp$, $E$, $G$, $B\flat$ is represented by a golden rectangle on the type ${\rm 2^*}$ and type ${\rm 3^*}$. }
\label{Bach12}
\end{figure}

\begin{figure}[H]
\centering
{%
\resizebox*{16cm}{!}{\includegraphics{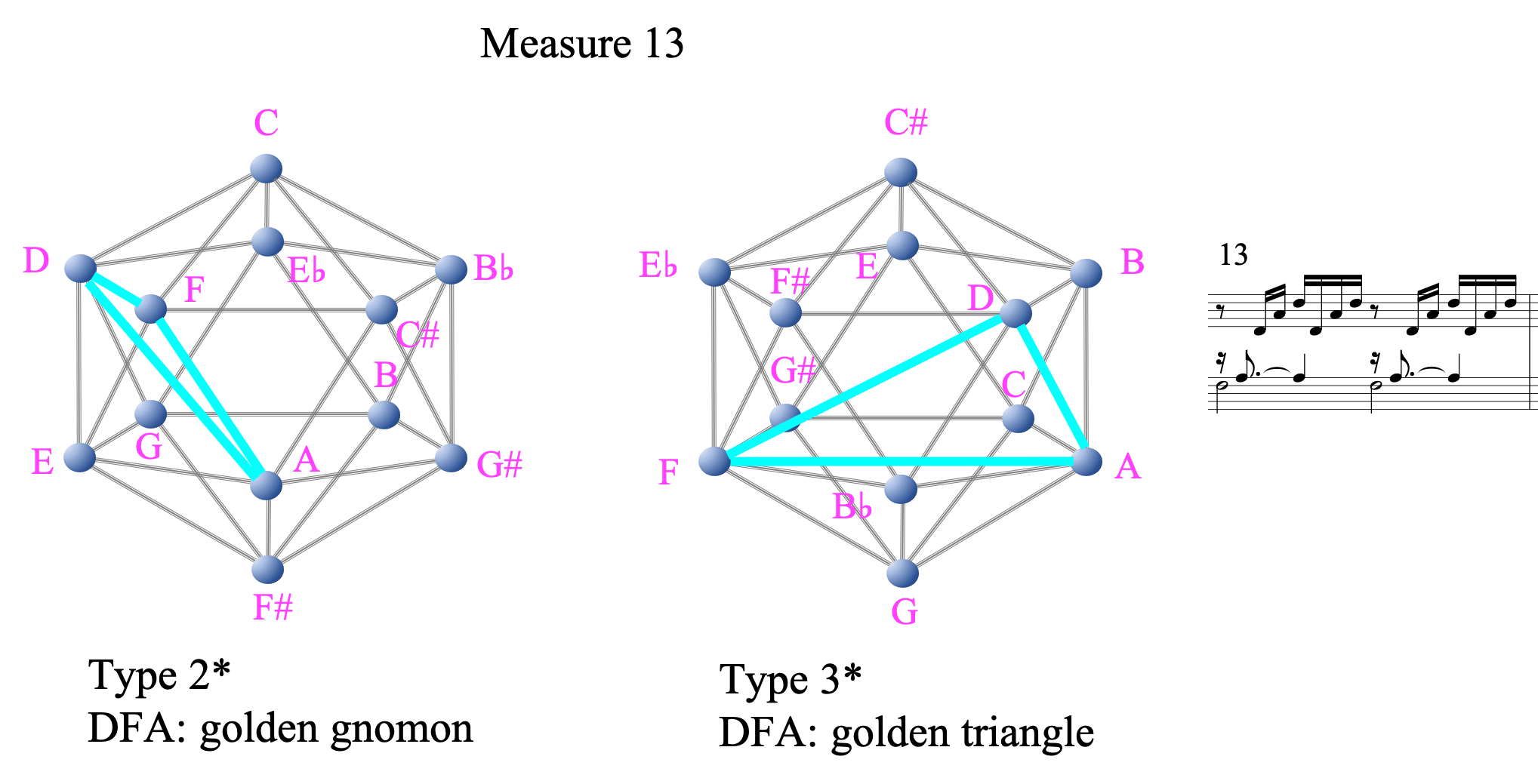}}}\hspace{5pt}
\caption{The golden analysis of the measure 13 of BWV 846 on the type ${\rm 2^*}$ and type ${\rm 3^*}$. $D$, $F$, $A$ is represented by a golden gnomon on the type ${\rm 2^*}$ and a golden triangle on the type ${\rm 3^*}$. }
\label{Bach13}
\end{figure}

\begin{figure}[H]
\centering
{%
\resizebox*{16cm}{!}{\includegraphics{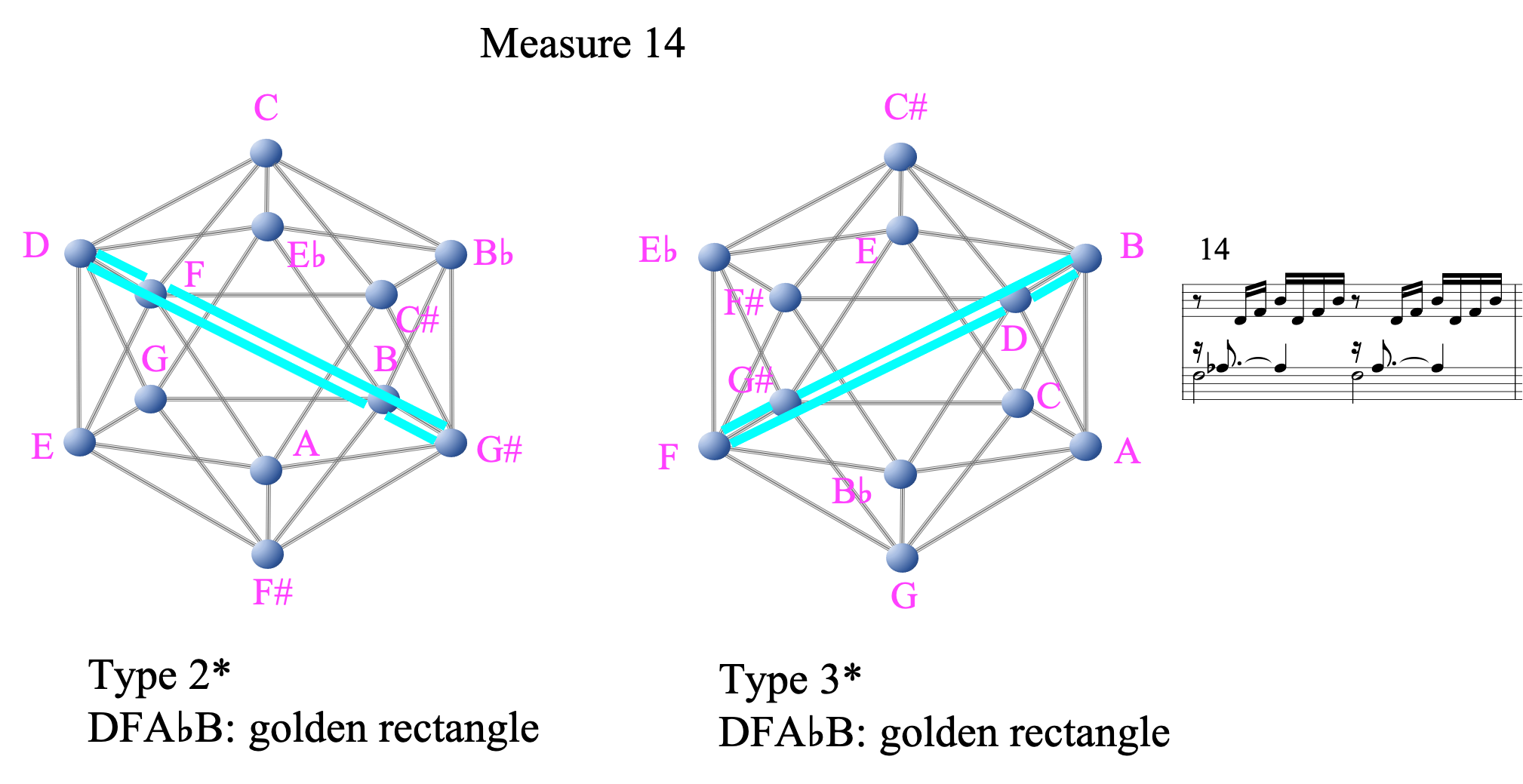}}}\hspace{5pt}
\caption{The golden analysis of the measure 14 of BWV 846 on the type ${\rm 2^*}$ and type ${\rm 3^*}$. $D$, $F$, $A\flat$, $B$ is represented by a golden rectangle on the type ${\rm 2^*}$ and type ${\rm 3^*}$. }
\label{Bach14}
\end{figure}

\begin{figure}[H]
\centering
{%
\resizebox*{16cm}{!}{\includegraphics{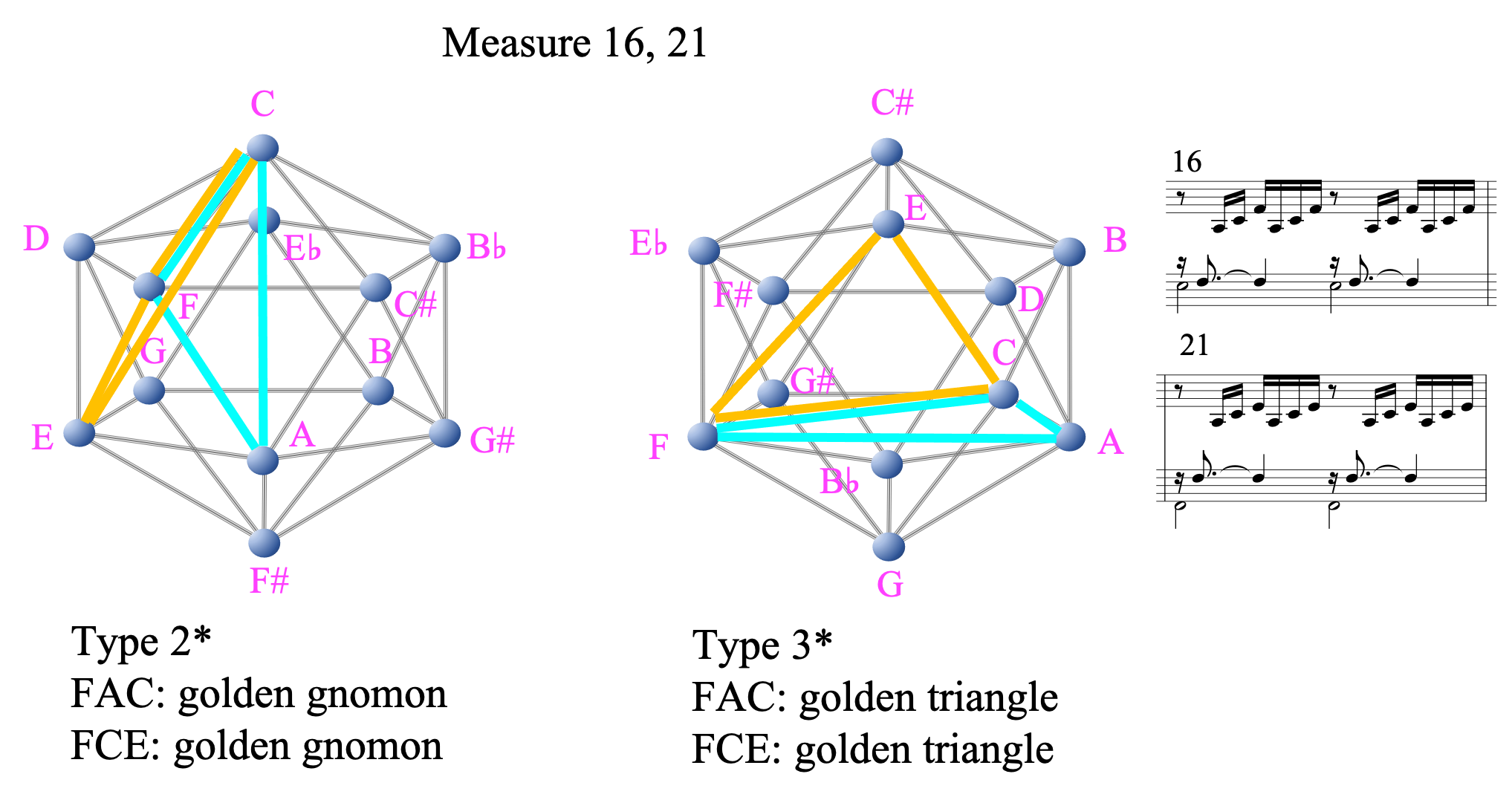}}}\hspace{5pt}
\caption{The golden analysis of the measure 16, 21 of BWV 846 on the type ${\rm 2^*}$ and type ${\rm 3^*}$. $F$, $A$, $C$, $E$ is represented by two golden gnomons ($F$, $A$, $C$ and $F$, $C$, $E$) on the type ${\rm 2^*}$ and two golden triangles ($F$, $A$, $C$ and $F$, $C$, $E$) on the type ${\rm 3^*}$. }
\label{Bach16}
\end{figure}

\begin{figure}[H]
\centering
{%
\resizebox*{16cm}{!}{\includegraphics{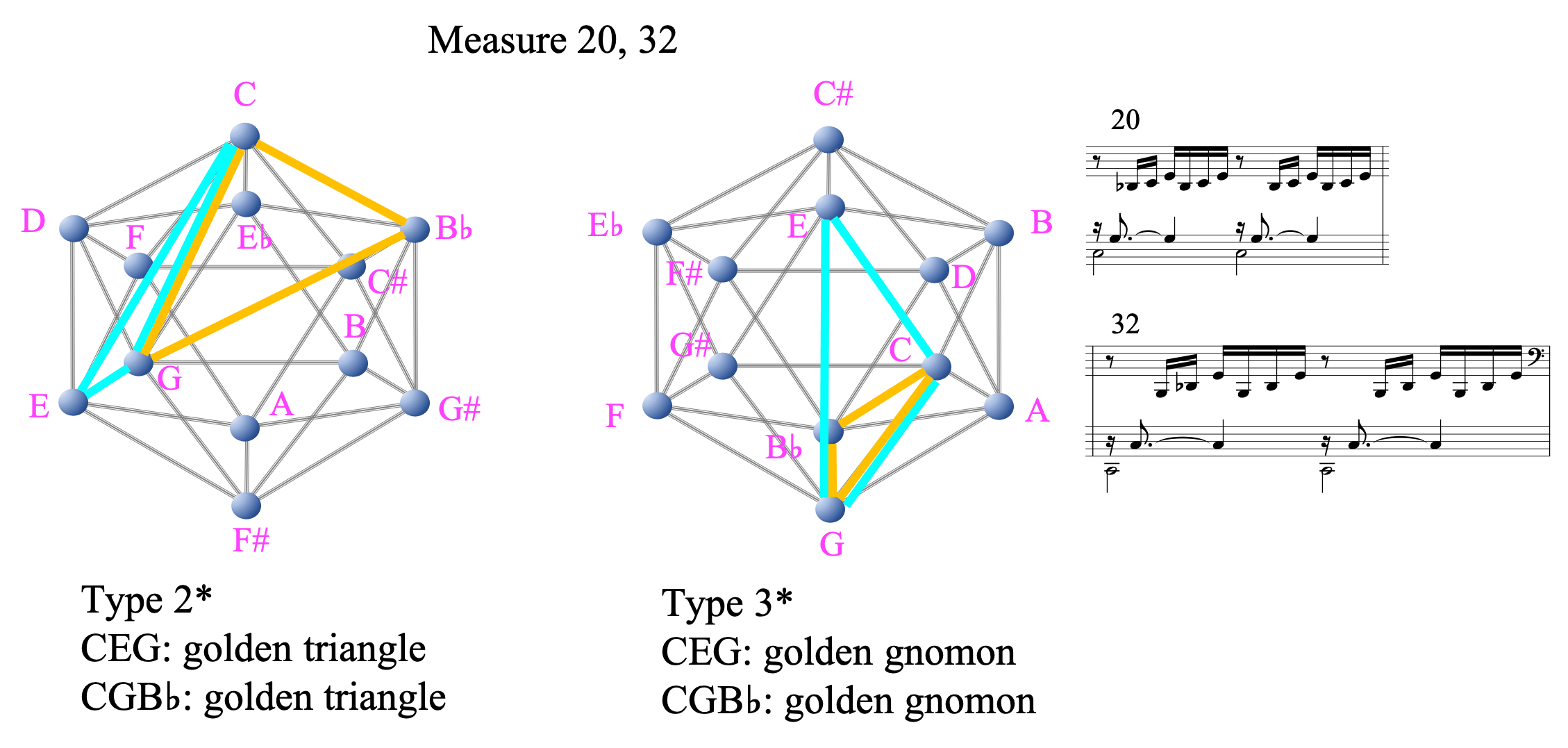}}}\hspace{5pt}
\caption{The golden analysis of the measure 20, 32 of BWV 846 on the type ${\rm 2^*}$ and type ${\rm 3^*}$. $C$, $E$, $G$, $B\flat$ is represented by two golden triangles ($C$, $E$, $G$ and $C$, $E$, $B\flat$) on the type ${\rm 2^*}$ and two golden gnomons ($C$, $E$, $G$ and $C$, $E$, $B\flat$) on the type ${\rm 3^*}$. }
\label{Bach20}
\end{figure}

\begin{figure}[H]
\centering
{%
\resizebox*{16cm}{!}{\includegraphics{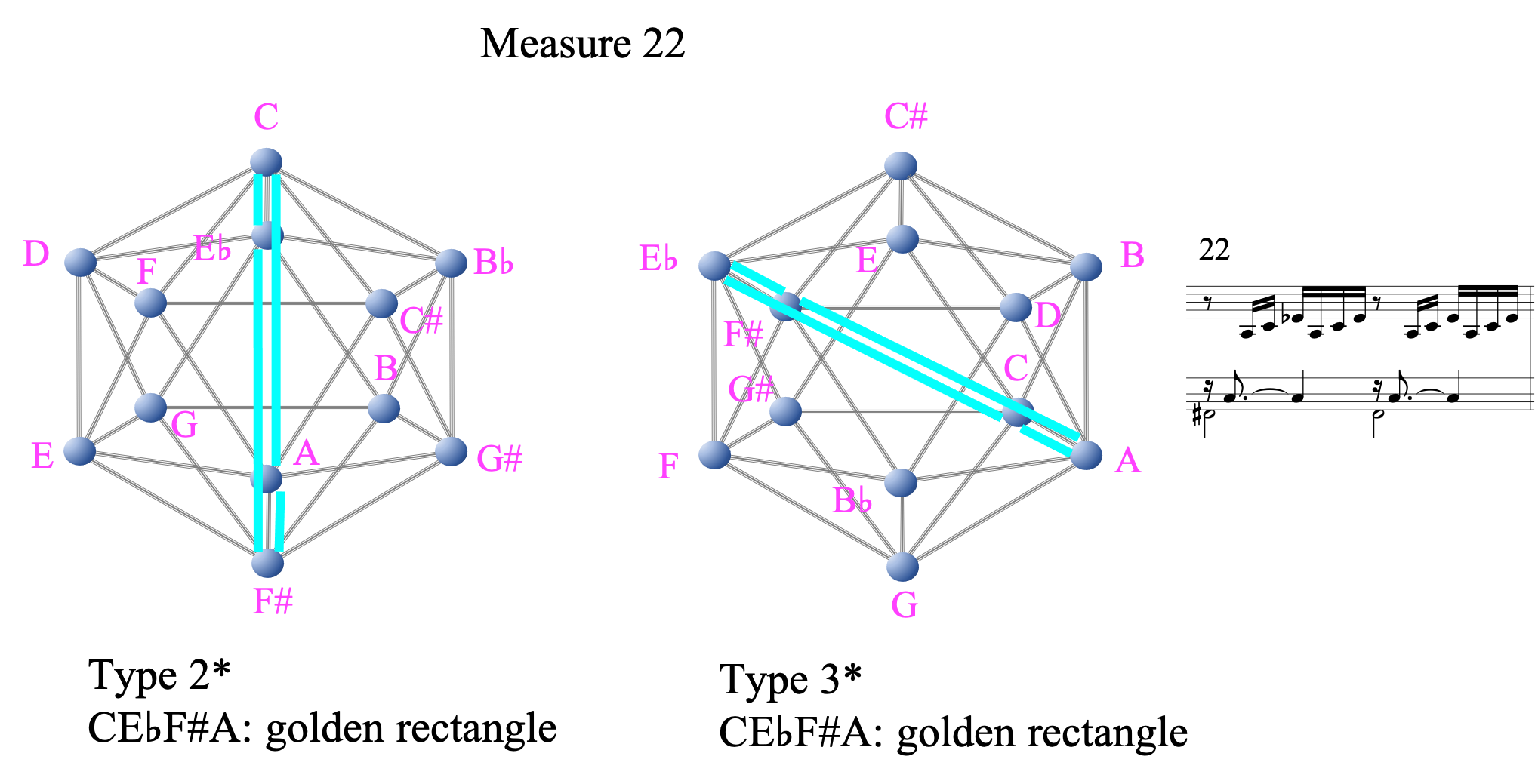}}}\hspace{5pt}
\caption{The golden analysis of the measure 22 of BWV 846 on the type ${\rm 2^*}$ and type ${\rm 3^*}$. $C$, $E\flat$, $F\sharp$, $A$ is represented by a golden rectangle on the type ${\rm 2^*}$ and type ${\rm 3^*}$. }
\label{Bach22}
\end{figure}

\begin{figure}[H]
\centering
{%
\resizebox*{16cm}{!}{\includegraphics{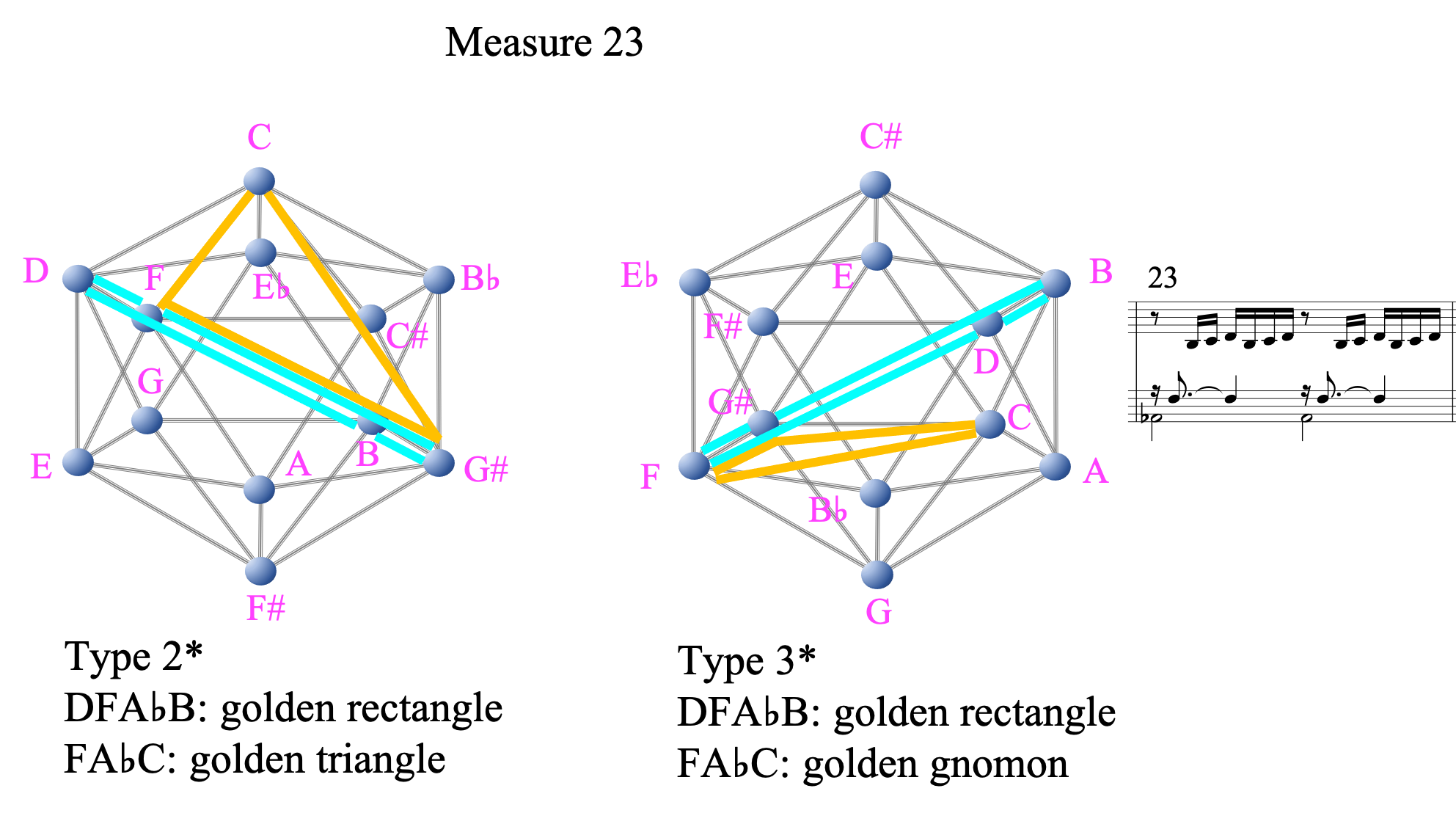}}}\hspace{5pt}
\caption{The golden analysis of the measure 23 of BWV 846 on the type ${\rm 2^*}$ and type ${\rm 3^*}$. $D$, $F$, $A\flat$, $B$, $C$ is represented by a golden rectangle and a golden triangle ($D$, $F$, $A\flat$, $B$ and $F$, $A\flat$, $C$) on the type ${\rm 2^*}$ and a golden rectangle and a golden gnomon ($D$, $F$, $A\flat$, $B$ and $F$, $A\flat$, $C$) on the type ${\rm 3^*}$. }
\label{Bach23}
\end{figure}

\begin{figure}[H]
\centering
{%
\resizebox*{16cm}{!}{\includegraphics{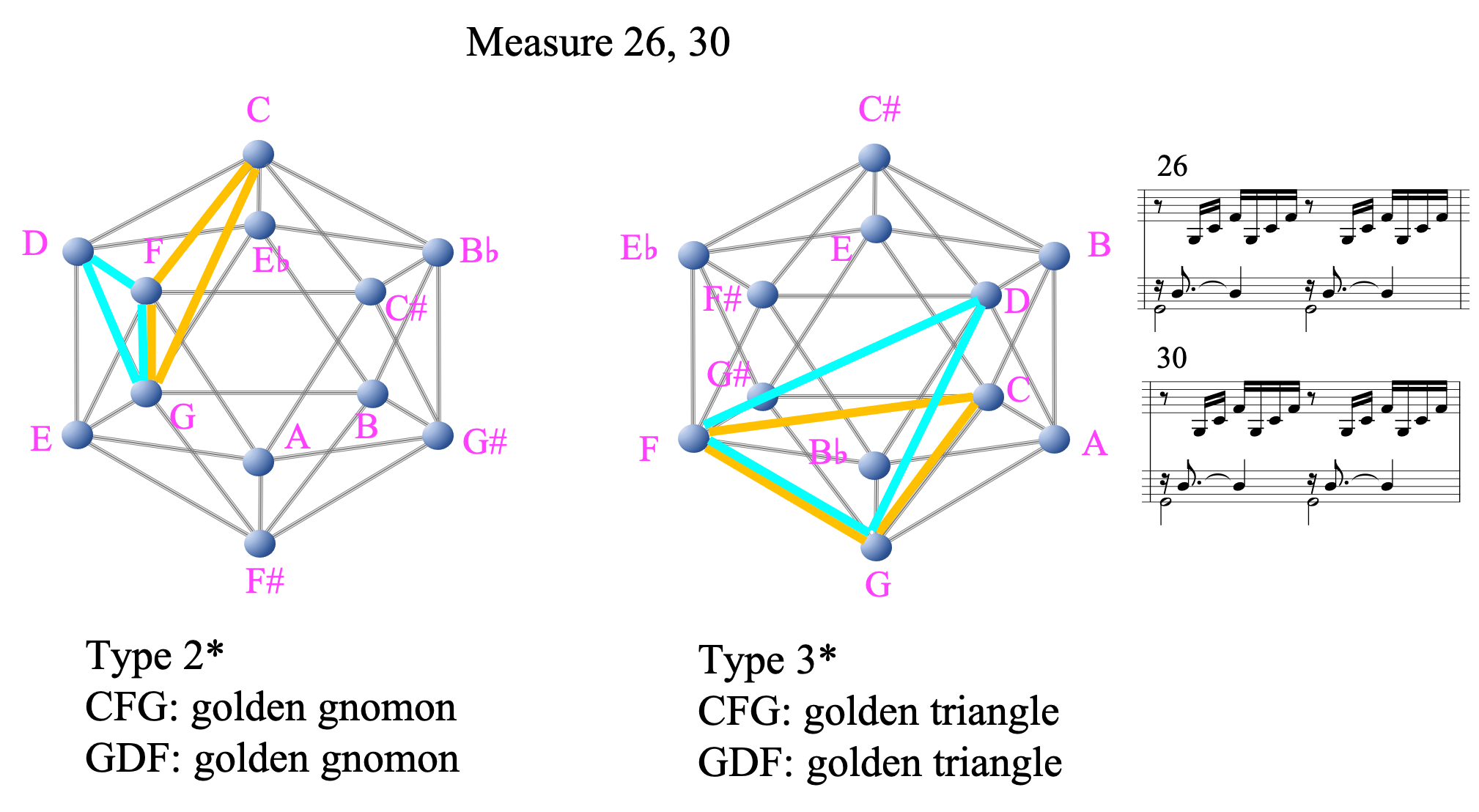}}}\hspace{5pt}
\caption{The golden analysis of the measure 26 of BWV 846 on the type ${\rm 2^*}$ and type ${\rm 3^*}$. $G$, $D$, $C$, $F$ is represented by two golden gnomons ($C$, $F$, $G$ and $G$, $D$, $F$) on the type ${\rm 2^*}$ and two golden triangles ($C$, $F$, $G$ and $G$, $D$, $F$) on the type ${\rm 3^*}$. }
\label{Bach26}
\end{figure}

\begin{figure}[H]
\centering
{%
\resizebox*{16cm}{!}{\includegraphics{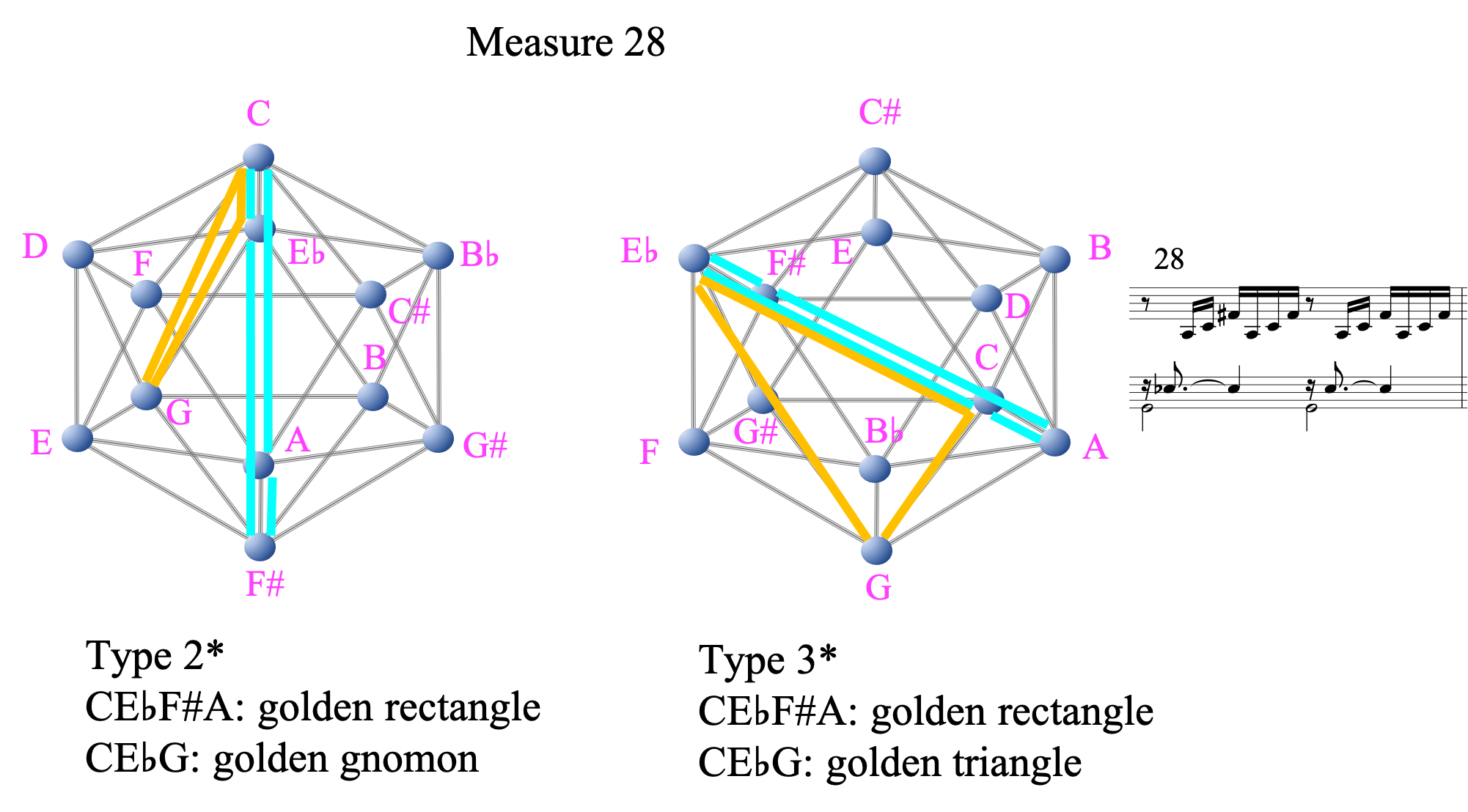}}}\hspace{5pt}
\caption{The golden analysis of the measure 28 of BWV 846 on the type ${\rm 2^*}$ and type ${\rm 3^*}$. $C$, $E\flat$, $F\sharp$, $G$, $A$ is represented by a golden rectangle and a golden gnomon ($C$, $E\flat$, $F\sharp$, $A$ and $C$, $E\flat$, $G$) on the type ${\rm 2^*}$ and a golden rectangle and a golden triangle ($C$, $E\flat$, $F\sharp$, $A$ and $C$, $E\flat$, $G$) on the type ${\rm 3^*}$. }
\label{Bach28}
\end{figure}

\begin{figure}[H]
\centering
{%
\resizebox*{16cm}{!}{\includegraphics{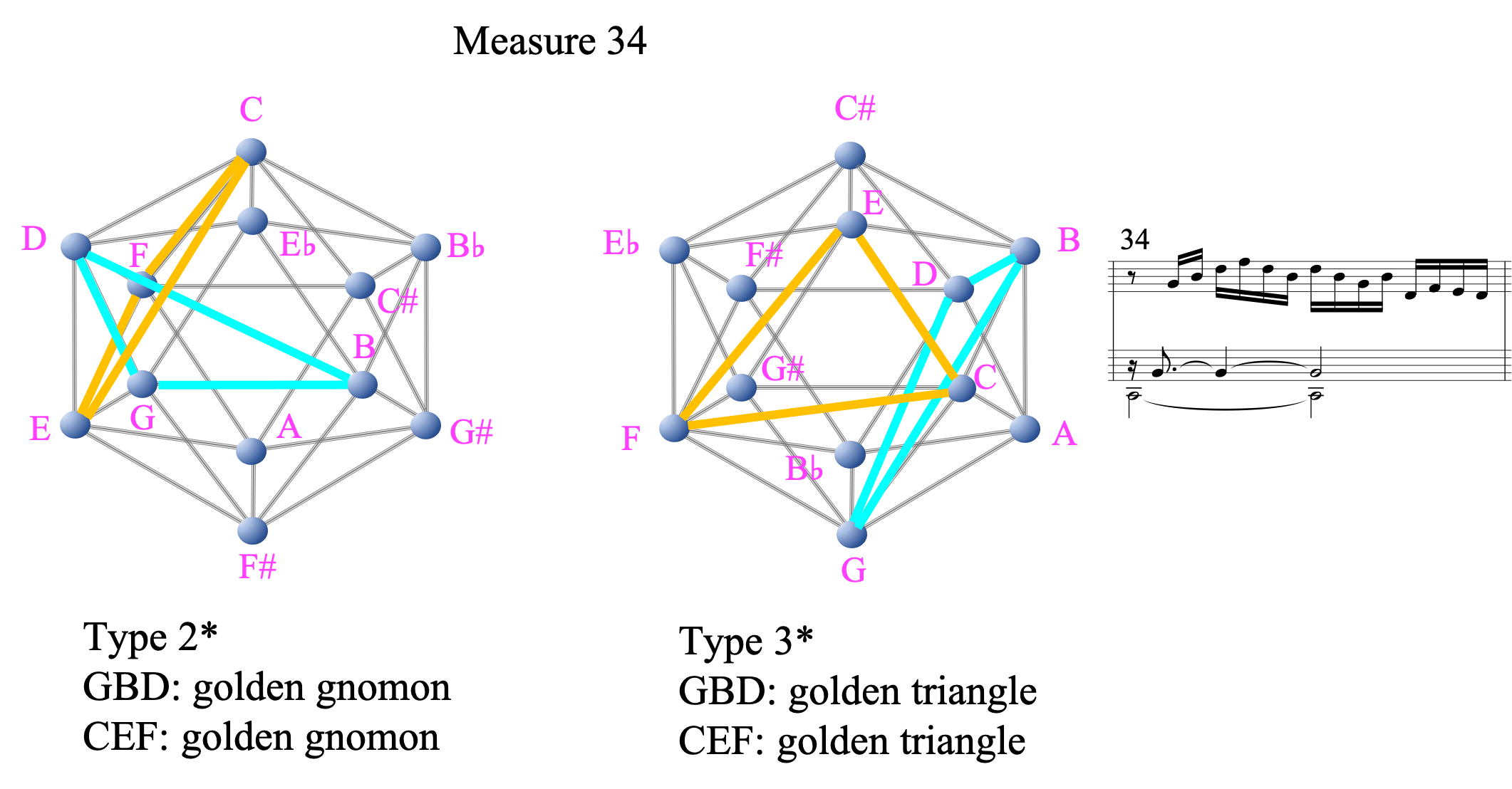}}}\hspace{5pt}
\caption{The golden analysis of the measure 34 of BWV 846 on the type ${\rm 2^*}$ and type ${\rm 3^*}$. $C$, $D$, $E$, $F$ $G$, $B$ is represented by two golden gnomons ($G$, $B$, $D$ and $C$, $E$, $F$) on the type ${\rm 2^*}$ and two golden triangles ($G$, $B$, $D$ and $C$, $E$, $F$) on the type ${\rm 3^*}$. }
\label{Bach34}
\end{figure}

\newpage
\section{Conclusion}
Many ways to analyze musical pieces have been proposed such as the chromatic circle, the circle of fifths, the Riemannian theory, the neo-Riemannian theory (\emph{Tonnetz}), and its various generalizations. In this paper, we proposed a new way of analyzing musical pieces by using the exceptional musical icosahedra introduced in our previous paper (Imai, Dellby, Tanaka, 2021).

First, we showed the relation between the exceptional musical icosahedra and the neo-Riemannian theory by using a new concept, golden neighborhood that characterizes neighboring golden triangles/gnomons of a given golden triangle or gnomon. Also, we found that the golden neighborhoods and the icosahedron symmetry relate any major/minor triad with any major/minor triad.

Second, we proposed a way to analyze musical pieces by the exceptional musical icosahedra: golden analysis. The golden analysis is an analysis of harmony by using golden decomposition that is defined as a decomposition of a given harmony into the minimum number of harmonies constructing the given harmony and represented by the golden figures (a golden triangle, a golden gnomon, or a golden rectangle). Because a golden triangle/gnomon on the type ${\rm 1^*}$ (type ${\rm 2^*}$) naturally corresponds to a golden gnomon/triangle on the type ${\rm 4^*}$ (type ${\rm 3^*}$), then, the golden analysis is dually constructed. We also introduced a new concept, golden singular. A harmony is golden singular if and only if the given harmony dose not have golden decompositions. While the dominant seventh chord is the only seventh chord that is golden singular in the ${\rm 1^*}$ and type ${\rm 4^*}$, the half-diminished chord is the only seventh chord that is golden singular in the ${\rm 2^*}$ and type ${\rm 3^*}$. Also, from the viewpoint of the golden decomposition, the minor major seventh chord can be regarded as a generalized major-minor dual of the augmented major seventh chord with some transpositions and interchanges. Furthermore, all the harmonies constructed by five or more tones is not golden singular.

Third, we apply the golden analysis to the famous prelude composed by Johann Sebastian Bach (BWV 846). We found 7 combinations of the golden figures (a golden triangle, a golden gnomon, two golden triangles, two golden gnomons, a golden rectangle, a golden rectangle and a golden triangle, a golden rectangle and a golden gnomon) on the type ${\rm 2^*}$ or the type ${\rm 3^*}$ can represent all the measures of the BWV 846. Because the BWV 846 includes the dominant seventh chord, type ${\rm 1^*}$ and the type ${\rm 4^*}$ are not appropriate as musical icosahedra that are applied to the golden analysis of the BWV 846.

Because the golden analysis characterizes many harmonies by the golden ratio, the golden analysis may be useful to visualize music beautifully. In addition to the BWV 846, there may be more musical pieces that can be visualized by combinations of figures characterized by the golden ratio. This paper should be regarded as the first step in making a theory of analyzing musical pieces by using musical icosahedra.

McCartin, B. J. (2012). A Geometric Demonstration of the Unique Intervallic Multiplicity Property of the Diatonic Musical Scale. \emph{International Mathematical Forum}. 7(57). 2815-2825.

Headlam, D. Hasegawa, R. Paul L., and Perle G. (2013). Twelve-note composition. \emph{Grove Music Online}.

Klumpenhouwer, H. (1994). Some Remarks on the Use of Riemann Transformations. \emph{Music Theory Online 0.9}.

Cohn, R. (1998). An Introduction to Neo-Riemannian Theory: A Survey and Historical Perspective. \emph{Journal of Music Theory}. 42(2): 167–180.

Cohn, R. (1997). Neo-Riemannian Operations, Parsimonious Trichords, and Their "Tonnetz" Representations, \emph{Journal of Music Theory}. 41(1). 1-66.

Euler, L. (1739). \emph{Tentamen novae theoriae musicae ex certissismis harmoniae principiis dilucide expositae}. Saint Petersburg Academy.

Cappuzo, G. (2004). Neo-Riemannian Theory and the Analysis of Pop-Rock Music. \emph{Music Theory Spectrum}. 26(2). 177–199.

Childs A. P. (1998). Moving beyond Neo-Riemannian Triads: Exploring a Transformational Model for Seventh Chords. \emph{Journal of Music Theory}. 42(2). 181-193.

Popoff, A. (2013). Building generalized neo-Riemannian groups of musical transformations as extensions. \emph{Journal of Mathematics and Music}. 7(1). 55-72.

Martin, N. (2008). The Tristan Chord Resolved. ``Intersections Canadian Journal of Music Revue canadienne de musique". 28(2). 6–30.

Benward, B. and Saker, M. (2009). \emph{Music IN THEORY AND PRACTICE}. New York: McGraw-Hill.

Imai, Y., Dellby, S. C., and Tanaka, N. (2021). General Theory of Music by Icosahedron 1: A bridge between "artificial" scales and "natural" scales, Duality between chromatic scale and Pythagorean chain, and Golden Major Minor Self-Duality. \emph{arXiv}: 2103.10272v3.

Lehman, F. (2014). Film music and Neo-Riemannian theory. \emph{Oxford Handbooks Online}.

Rusch, R. (2013). Schenkerian Theory, Neo-Riemannian Theory and Late Schubert: A Lesson from Tovey. \emph{Journal of the Society for Musicology in Ireland}. 8. 3-20.

Clough, J. (2002). Diatonic Trichords in Two Pieces from Kurt\'ag's Kafka-Fragmente: A Neo-Riemannian Approach. \emph{Studia Musicologica Academiae Scientiarum Hungaricae}. 43(3-4). 333-344.

Briginshaw, Sara B.P. (2012). A Neo-Riemannian Approach to Jazz Analysis. \emph{Nota Bene: Canadian Undergraduate Journal of Musicology}. 5(1). Article 5.

Catanzaro, M. J. (2011). Generalized \emph{Tonnetze}. \emph{Journal of Mathematics and Music}. 5(2). 117-139.

Mohanty, V. (2020). A 5-dimensional Tonnetz for nearly symmetric hexachords, \emph{Journal of Mathematics and Music}. 5(2). Latest Articles.

Baroin, G. (2011). Mathematics and Computation in Music, The Planet-4D Model: An Original Hypersymmetric Music Space Based on Graph Theory. In Agon, C.; Andreatta, M.; Assayag, G.; Amiot, E.; Bresson, J.; Mandereau, J. (eds.). \emph{Mathematics and Computation in Music}. Lecture Notes in Computer Science. Berlin, Heidelberg: Springer. 326–329.

Amiot, E. (2013). The Torii of Phases, Mathematics and Computation in Music. In Yust, J.; Wild, J.; Burgoyne, J. A. (eds.). \emph{Mathematics and Computation in Music}. Lecture Notes in Computer Science. Berlin, Heidelberg: Springer Berlin Heidelberg. 1–18.

Wason, R. W. (1982). \emph{Viennese harmonic theory from Albrchetsberger to Schenker and Schoenberg}, Ann Arbor: University Microfilms International.

\end{document}